\documentclass{aastex631}

\usepackage[figuresright]{rotating}
\shorttitle{Peptide-like molecules in Sgr B2}
\shortauthors{Zheng et al.}
\graphicspath{{./}{figures/}}

\begin{document}

\title{Mapping Observations of Peptide-like molecules around Sagittarius B2\footnote{Released on March, 1st, 2021}}



\author{Siqi Zheng}\thanks{E-mail: zhengsq@shao.ac.cn}
\affiliation{Shanghai Astronomical Observatory, Chinese Academy of Sciences\\
No. 80 Nandan Road  \\
Shanghai, 200030, China}
\affiliation{Key Laboratory of Radio Astronomy, Chinese Academy of Sciences\\
Nanjing 210033, China}
\affiliation{School of Astronomy and Space Sciences, University of Chinese Academy of Sciences \\
No. 19A Yuquan Road \\
Beijing 100049, People's Republic of China}

\author{Juan Li}\thanks{E-mail: lijuan@shao.ac.cn}
\affiliation{Shanghai Astronomical Observatory, Chinese Academy of Sciences\\
No. 80 Nandan Road  \\
Shanghai, 200030, China}
\affiliation{Key Laboratory of Radio Astronomy, Chinese Academy of Sciences\\
Nanjing 210033, China}

\author{Junzhi Wang}\thanks{E-mail: junzhiwang@gxu.edu.cn}
\affiliation{Guangxi Key Laboratory for Relativistic Astrophysics, Department of Physics, Guangxi University\\
 Nanning 530004, PR China}

\author{Yao Wang}
\affiliation{ Purple Mountain Observatory, Chinese Academy of Sciences, 10 Yuanhua Road, Nanjing 210023, People's Republic of China 
}

\author{Feng Gao}
\affiliation{ Hamburger Sternwarte, Universit{\"a}t Hamburg\\
Gojenbergsweg 112, D-21029, Hamburg, Germany
}
\affiliation{Max Planck Institute for Extraterrestrial Physics (MPE), Giessenbachstr. 1, D-85748 Garching, Germany}

\author{Donghui Quan}
\affiliation{ Xinjiang Astronomical Observatory, Chinese Academy of Sciences, 150 Science 1-Street, Urumqi 830011, People's Republic of China \\
}
\affiliation{Research Center for Intelligent Computing Platforms, Zhejiang Laboratory, Hangzhou 311100, China}

\author{Fujun Du}
\affiliation{ Purple Mountain Observatory, Chinese Academy of Sciences, 10 Yuanhua Road, Nanjing 210023, People's Republic of China 
}
\affiliation{ School of Astronomy and Space Science, University of Science and Technology of China, Hefei 230026, People's Republic of China}

\author{Yajun Wu}
\affiliation{Shanghai Astronomical Observatory, Chinese Academy of Sciences\\
No. 80 Nandan Road  \\
Shanghai, 200030, China}
\affiliation{Key Laboratory of Radio Astronomy, Chinese Academy of Sciences\\
Nanjing 210033, China}

\author{Edwin Bergin}
\affiliation{ Department of Astronomy, University of Michigan, Ann Arbor, MI 48109, USA
}

\author{Yuqiang Li}
\affiliation{Shanghai Astronomical Observatory, Chinese Academy of Sciences\\
No. 80 Nandan Road  \\
Shanghai, 200030, China}
\affiliation{Key Laboratory of Radio Astronomy, Chinese Academy of Sciences\\
Nanjing 210033, China}
\affiliation{School of Astronomy and Space Sciences, University of Chinese Academy of Sciences \\
No. 19A Yuquan Road \\
Beijing 100049, People's Republic of China}




\begin{abstract}

Peptide-like molecule, which has a close connection with the origin of life, has been detected in universe. 
Mapping observations of HCONH$_2$ and CH$_3$CONH$_2$, two simplest peptide-like molecules, are performed towards Sagittarius B2 (Sgr B2) complex with the IRAM 30m telescope.
Seven transitions of HCONH$_2$ and five transitions of CH$_3$CONH$_2$ are used in analysis. 
The spatial distribution of excitation temperature and column density of HCONH$_2$ in the molecular envelope of Sgr B2 are obtained by the rotation diagrams.  Assuming the same excitation temperature as HCONH$_2$, the column densities of CH$_3$CONH$_2$ are also calculated. 
The results show that excitation temperature ranges from 6 K to 46 K in the molecular envelope of Sgr B2. 
The abundance ratio between HCONH$_2$ and CH$_3$CONH$_2$ are calculated to explore the relationship among them, as well as HNCO mentioned in our pervious research. 
The abundance ratio of CH$_3$CONH$_2$/HCONH$_2$ varies from 10\% to 20\%, while that of HCONH$_2$/HNCO ranges from 1.5\% to 10\%. CH$_3$CONH$_2$ is enhanced with respect to HCONH$_2$ in the northwest region of Sgr B2.
One transition of H$^{13}$CONH$_2$ is detected toward 12 positions of Sgr B2, from which a $^{12}C$/$^{13}C$ ratio of 28.7 is obtained. 
A time-dependent chemical model with a short duration of X-ray burst is used to explain the observed abundances of HCONH$_2$ and CH$_3$CONH$_2$, with the best fitting result at $T_{\mathrm{dust}}$ = 53-56 K. 
More chemical reactions are required to be included into the model since the modeled abundance is lower than the observed one at the observed $T_{\mathrm{dust}}$. 

\end{abstract}

 
\section{Introduction} \label{sec:intro}
Peptide bond [-NH-C(O)-] is the element that links two amino acids together to form proteins. As a result, peptide-like molecules, the molecules with a structure like the peptide bond, can give us some hints into the origin of life in universe. 
Formamide (HCONH$_2$) and acetamide (CH$_3$CONH$_2$) are the smallest two peptide-like molecules.
HCONH$_2$ has been detected in many high-mass star formation regions, comets, low-mass star formation regions and translucent cloud \citep[e.g.][]{2015MNRAS.449.2438L,2021A&A...653A.129C}. It is regarded as the precursor of many other prebiotic molecules \citep{C2CS35066A}. CH$_3$CONH$_2$ has been detected in  high-mass star formation regions \citep[e.g.][]{2006ApJ...643L..25H,2011ApJ...743...60H,2021A&A...653A.129C}, intermediate-mass protostar Serpens SMM1-a \citep{2022ESC.....6..455L}, and tentatively detected in low-mass protostar IRAS 16293-2422 \citep{2018MNRAS.480.3628L}. 
CH$_3$CONH$_2$ was only detected toward several sources, which was much less detected than that of HCONH$_2$.

Many chemical models have been proposed based on experiments and theoretical calculations to explain their abundance observed in ISM, including gas-phase reactions and grain surface reactions. 
The formation of HCONH$_2$ is still under discussion (see \citet{2019ESC.....3.2122L} for detail). 
\cite{2015MNRAS.453L..31B} suggested that H$_2$CO + NH$_2$ $\longrightarrow$ HCONH$_2$ + H was an effective way for HCONH$_2$ to form in gas-phase reaction according to the results of their electronic structure and kinetice calculation. \cite{2017MNRAS.468L...1S} calculated the deuteration rate of HCONH$_2$ in this reaction. The results of computations were in good agreement with the observation in the hot corino of IRAS16293-2422 B. 
This reaction has also been used to explain the observed spatial segregtion of HCONH$_2$ with respect to other species in shock region L1157-B1 by \citet{2017A&A...605L...3C}.
A newly proposed grain surface reaction is the reaction between CN radial and the water molecule of the ice mantle \citep{2018ESC.....2..720R}. 
The hydrogenation of HNCO was ruled out by experiments \citep{2015A&A...576A..91N}, but \cite{doi:10.1021/jacs.9b04491} proposed a model that link HNCO and HCONH$_2$ with a double-cycle based on their experiment results.
For CH$_3$CONH$_2$, many gas-phase reactions and grain-surface reactions have also been proposed \citep[e.g.][]{2007A&A...474..521Q,2018MNRAS.474.2796Q}.
A constant ratio between HCONH$_2$ and CH$_3$CONH$_2$ in various physical environments indicates that they form at an early state of star formation on the interstellar dust grains \citep{2022ESC.....6..455L}. 
Which kind of reactions are most important for reproducing the abundance of CH$_3$CONH$_2$ and HCONH$_2$ in different environment  need to be constrained by more observation results.

Sagittarius B2 (hereafter Sgr B2) is an ideal place to study the chemical origin and the relationship between HCONH$_2$ and CH$_3$CONH$_2$, where many complex molecules have been detected to have an extended distribution and strong emission there \citep[e.g.][]{2008MNRAS.386..117J, 2017ApJ...849..115L}. 
 It consists of three parts: a low density envelop, a moderate density region and molecular hotspots where the high-mass stars are forming \citep[][]{1993A&A...276..445H}.
The two major high-mass star formation regions with high chemical richness are called Sgr B2(N) and Sgr B2(M). Both HCONH$_2$ and CH$_3$CONH$_2$ have been detected in Sgr B2(N) \citep{2006ApJ...643L..25H,2011ApJ...743...60H}. Later, in the hot cores of it, such as Sgr B2 (N1), Sgr B2 (N2), Sgr B2 (N3), Sgr B2(N4) and Sgr B2(N5), these two molecules has also been detected and analysed \citep[e.g.][]{2017A&A...604A..60B,2019A&A...628A..10B,2021ApJ...919....4L}. For Sgr B2(M), only HCONH$_2$ has been detected, while CH$_3$CONH$_2$ has not been detected \citep[][]{2013A&A...559A..47B}. The molecular envelop has an average H$_2$ density of $\sim$ 10$^3$ cm$^{-3}$, and H$_2$ column density of $\sim$ 10$^{23}$ cm$^{-2}$ \citep{1990ApJ...350..186G}. The size of it is 38 pc (15.4 \arcmin) in diameter \citep{1975ApJ...201..352S}. The spatial distribution of different molecules here was found to be significantly different \citep[e.g.][]{2008MNRAS.386..117J, 2011MNRAS.411.2293J}. In the molecular envelop of Sgr B2, G+0.693-0.27 (hereafter G+0.693) is a quiescent giant molecular cloud with great interest. This is a source with low excitation temperature and high levels of chemical richness \citep{2008ApJ...672..352R,2018MNRAS.478.2962Z}, while there is no sign of active star formation \citep{2020MNRAS.497.4896Z}. Recently, \cite{2023MNRAS.523.1448Z} detected both molecules in G+0.693. 
While the observation towards the hot cores and cold envelop of Sgr B2 are carried out, all of the observations mentioned above are performed toward a single position. 
The simultaneous mapping observation of both molecules can not only avoid the uncertainties introduced by absolute flux calibration and pointing, but also
tell us how the abundance ratio varies with the environment. The spatial variation of the abundance ratio indicates the possible mechanisms that dominate the chemical networks.
Besides, the physical conditions of the extended molecular envelope of Sgr B2, which remain unclear now, could be obtained with multiple transitions of these molecules. 

In this paper, we present mapping observation results of HCONH$_2$ and CH$_3$CONH$_2$ with the IRAM 30m toward Sgr B2 complex. We study the spatial distribution and the abundance ratio of HCONH$_2$ and CH$_3$CONH$_2$ in Sgr B2, and the possible chemical paths would be proposed according to the abundance ratio. In Section \ref{sec:observation and data reduction}, we introduce the observation and data reduction. In Section \ref{sec:results}, we present the results of our observation. In Section \ref{sec:discussion}, we make a scientific discussion about the results. The conclusions are summarised in Section \ref{sec:summary}.

\section{Observation and data reduction} \label{sec:observation and data reduction}


We have performed mapping observation toward Sgr B2 with the IRAM 30m telescope on Pico Veleta Spain (project 170-18) in 2019 May. The Eight MIxer Receiver and FFTs in FTS200 mode were used in position-switching mode. The observation were performed at the 3 mm band, covering a frequency range of 82.3-90 GHz. The channel spacing is 0.195 MHz, which corresponds to the velocity resolution of 0.64 km s$^{-1}$ at 84 GHz. Pointing was checked every $\sim$2 h on 1757-240, and the focus was calibrated on 1757-240 before the observation. The integration time ranges from 24 minutes to 98 minutes for different positions, with a typical system temperature of $\sim$110 K, which gives a 1$\sigma$ rms of 4-8 mK in $T_{\rm A}^{\ast}$ derived with the line free channels. The calibration uncertainty is within 20\%.

The observing center is Sgr B2(N) ($\alpha_{J2000}=17^h47^m20^s.0$,$\delta_{J2000}=-28^{\circ}22\arcmin 19.0\arcsec$), with a sampling interval of 30\arcsec and a beam size of 30\arcsec. The off position of ($\delta \alpha$, $\delta \beta$)=(-752\arcsec, 342\arcsec) with respect to Sgr B2(N) was used \citep{2013A&A...559A..47B}. 
The data were reduced with CLASS software\footnote{\tt http://www.iram.fr/IRAMFR/GILDAS.}. The continuum emission was substracted via first-order polynomial fitting.
The antenna temperature ($T_{\rm A}^{\ast}$) was transformed into the main beam brightness temperature ($T_{\rm mb}$) using $T_{\rm mb}$=$T_{\rm A}^{\ast}\cdot F_{\rm eff}/B_{\rm eff}$, where the forward efficiency $F_{\rm eff}$ is 0.95 at 3 mm, and the beam efficiency $B_{\rm eff}$ is attained using Ruze's equation.
The spectra are smoothed to 1.35 km s$^{-1}$ to improve the signal-to-noise ratio.

\section{Results} \label{sec:results}
63 positions of Sgr B2 are observed to make the spatial distribution maps of target molecules. 
Due to the strong continuum and complex environment in Sgr B2(N) and Sgr B2(M),
it would be difficult to do baseline correction and to separate the absorption or emission that comes from hot cores or the cold envelope. Therefore, the spectra of molecules in both Sgr B2(N) and Sgr B2(M) are not considered in the further analysis. Among the transitions of HCONH$_2$ and CH$_3$CONH$_2$, the transitions with high upper state energies are only detected around the hot cores, while in the cold envelope only transitions with low upper state energies ($E_u < $ 40 K) can be detected. Besides, many transitions of the target molecules are blended with the transitions of other species due to the large line width of spectra in Sgr B2 \citep[e.g.][]{2013A&A...559A..47B}. Finally, seven transitions of HCONH$_2$ and five transitions of CH$_3$CONH$_2$ are selected because their emission is clean and strong enough to be detected in the cold envelop. One transition of H$^{13}$CONH$_2$ are also used for analysis. The information of these transitions is listed in Table \ref{transition}.
The spectroscopic entries from Cologne Database for Molecular Spectroscopy \citep{2005JMoSt.742..215M} are used. The transition data of CH$_3$CONH$_2$ was provided by \cite{2004JMoSp.227..115I}.

\subsection{HCONH$_2$} \label{subsec:hconh2}

\subsubsection{Distribution of HCONH$_2$}

\begin{figure}
    \centering
    \includegraphics[width=0.32\textwidth]{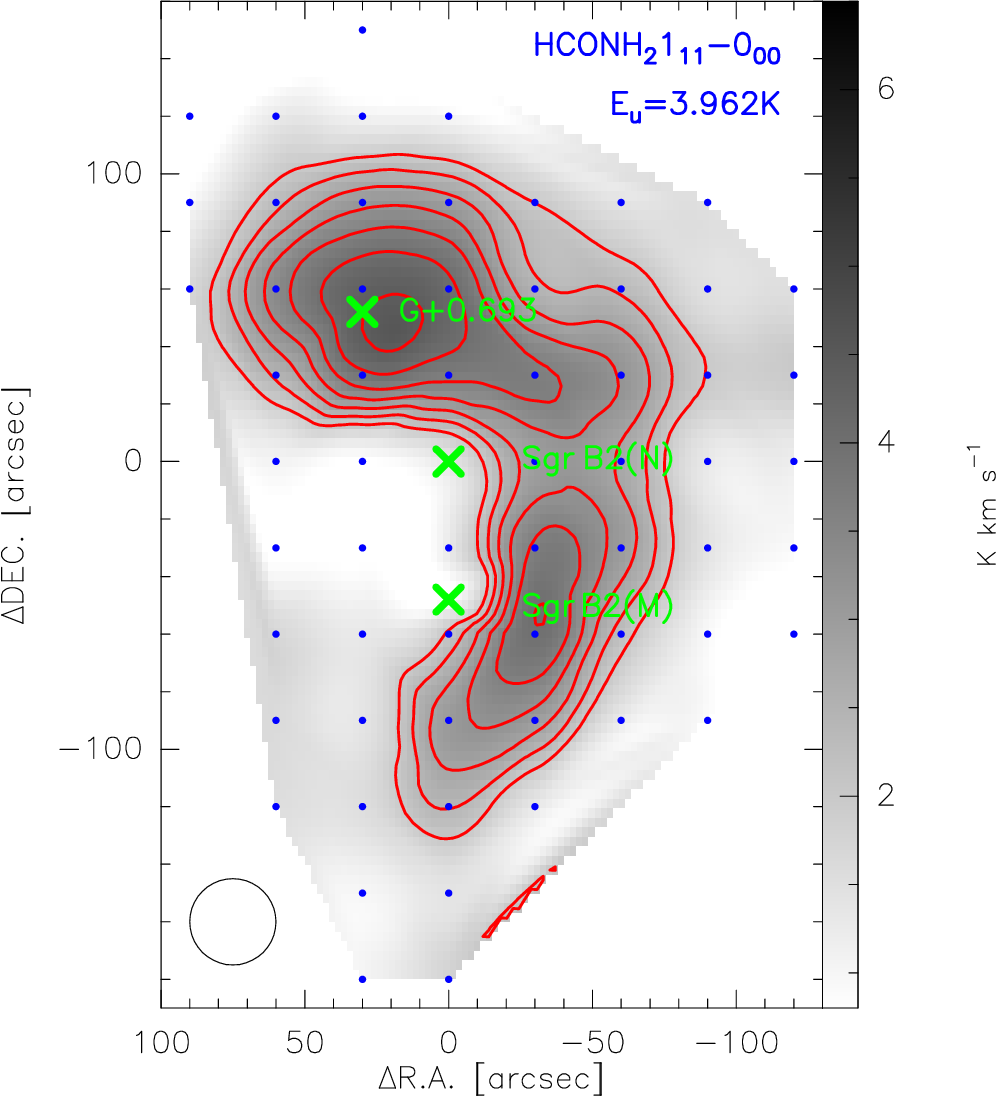}
    \includegraphics[width=0.32\textwidth]{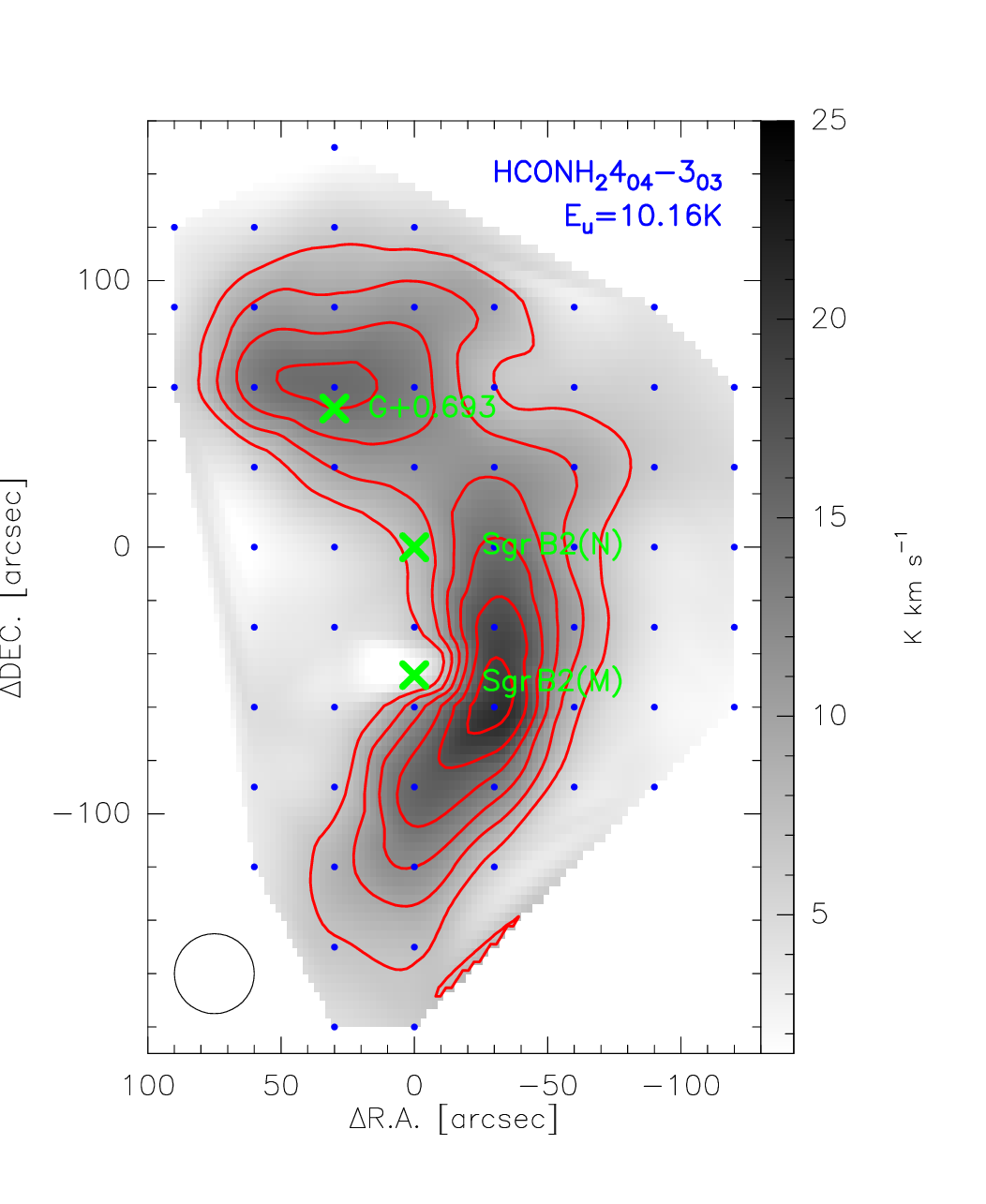}
    \includegraphics[width=0.32\textwidth]{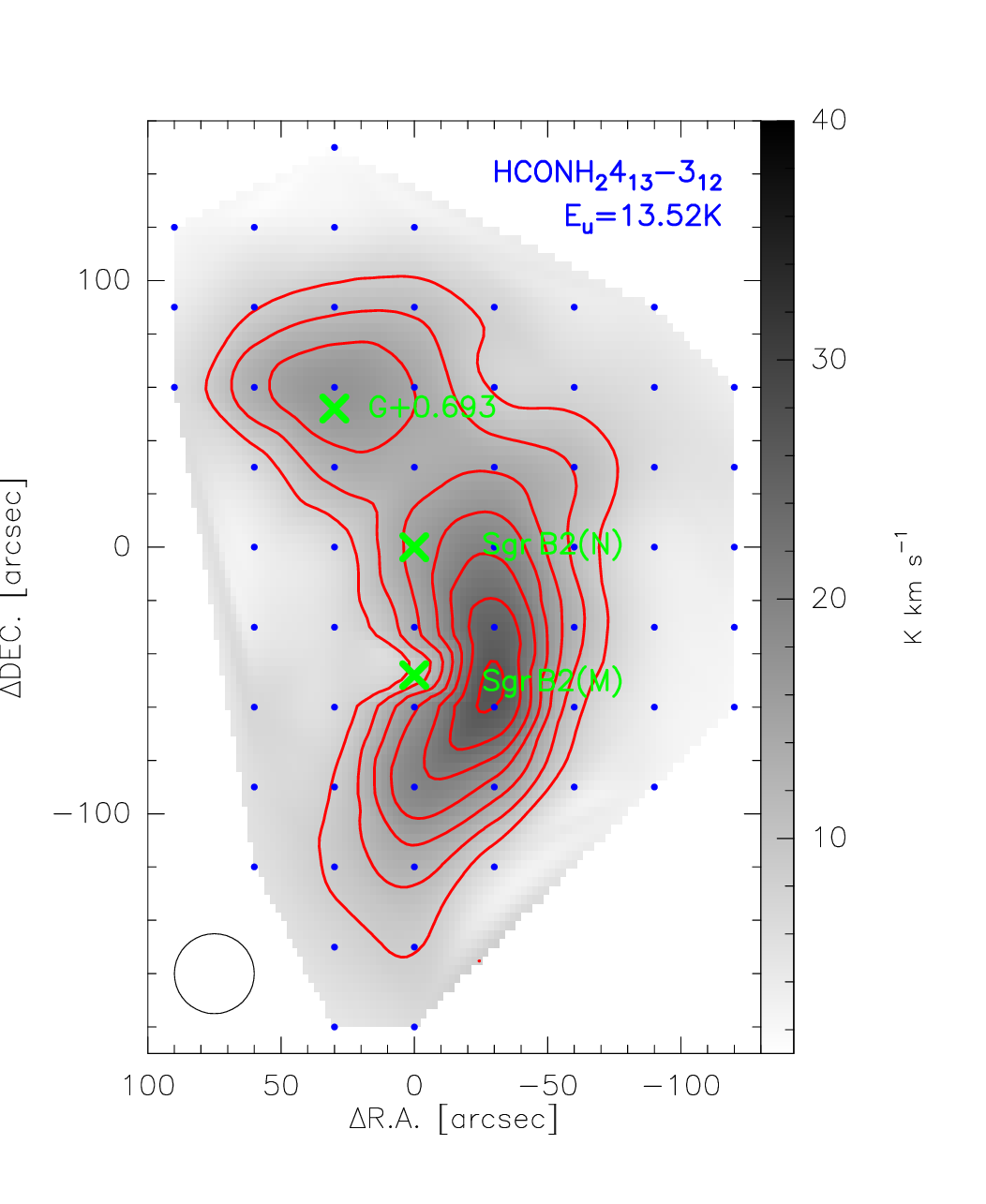}
    \includegraphics[width=0.32\textwidth]{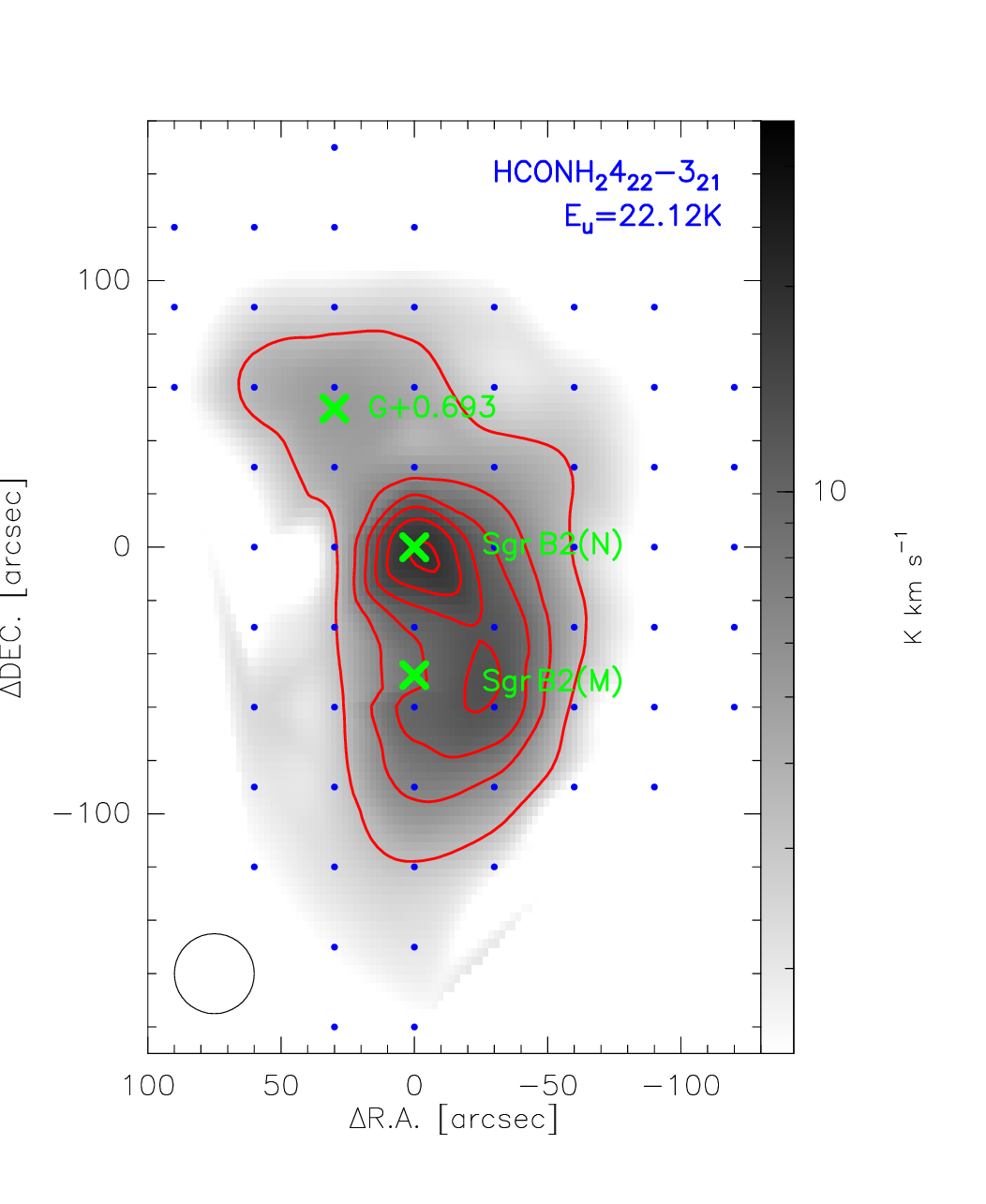}
    \includegraphics[width=0.32\textwidth]{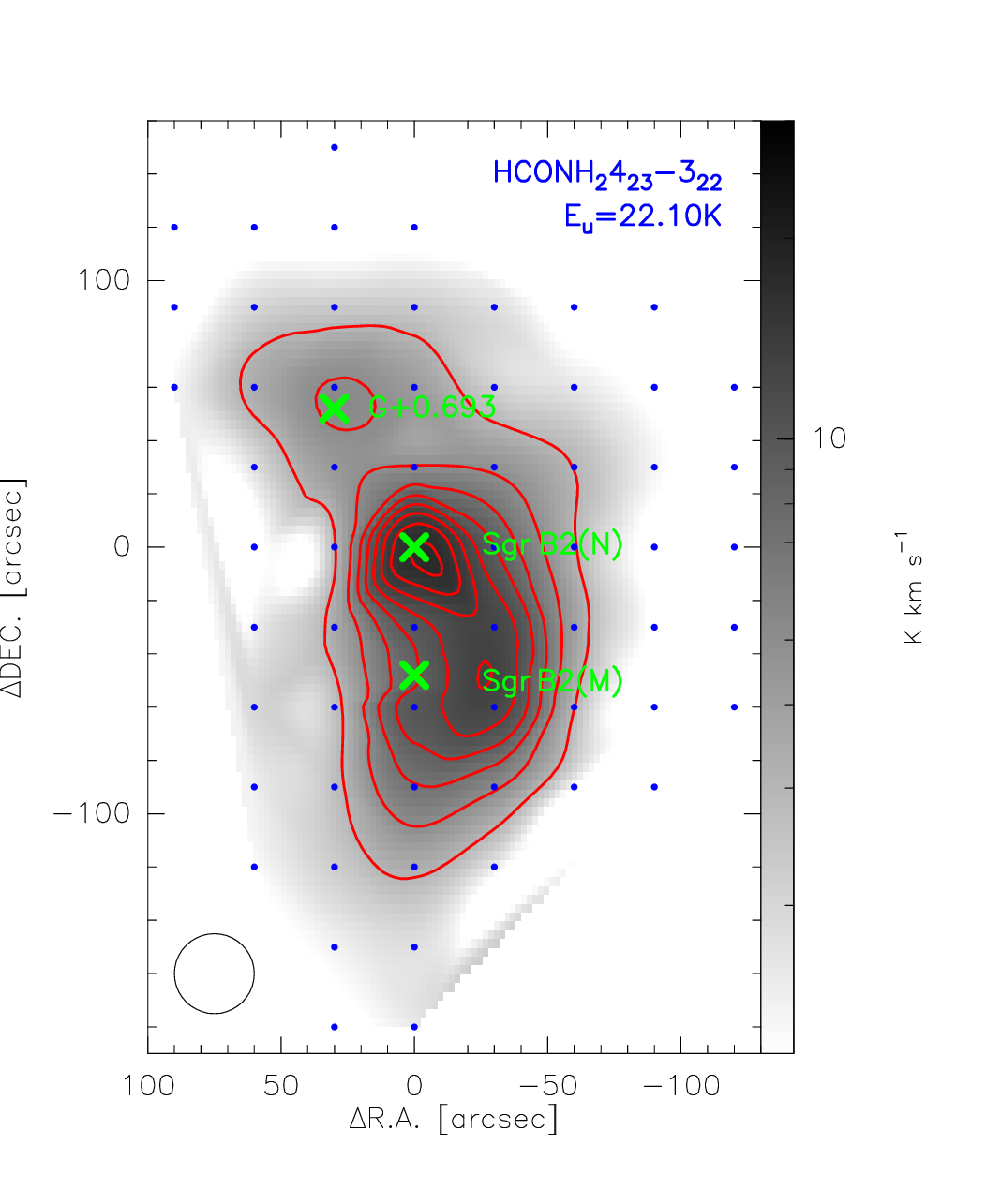}
    \includegraphics[width=0.32\textwidth]{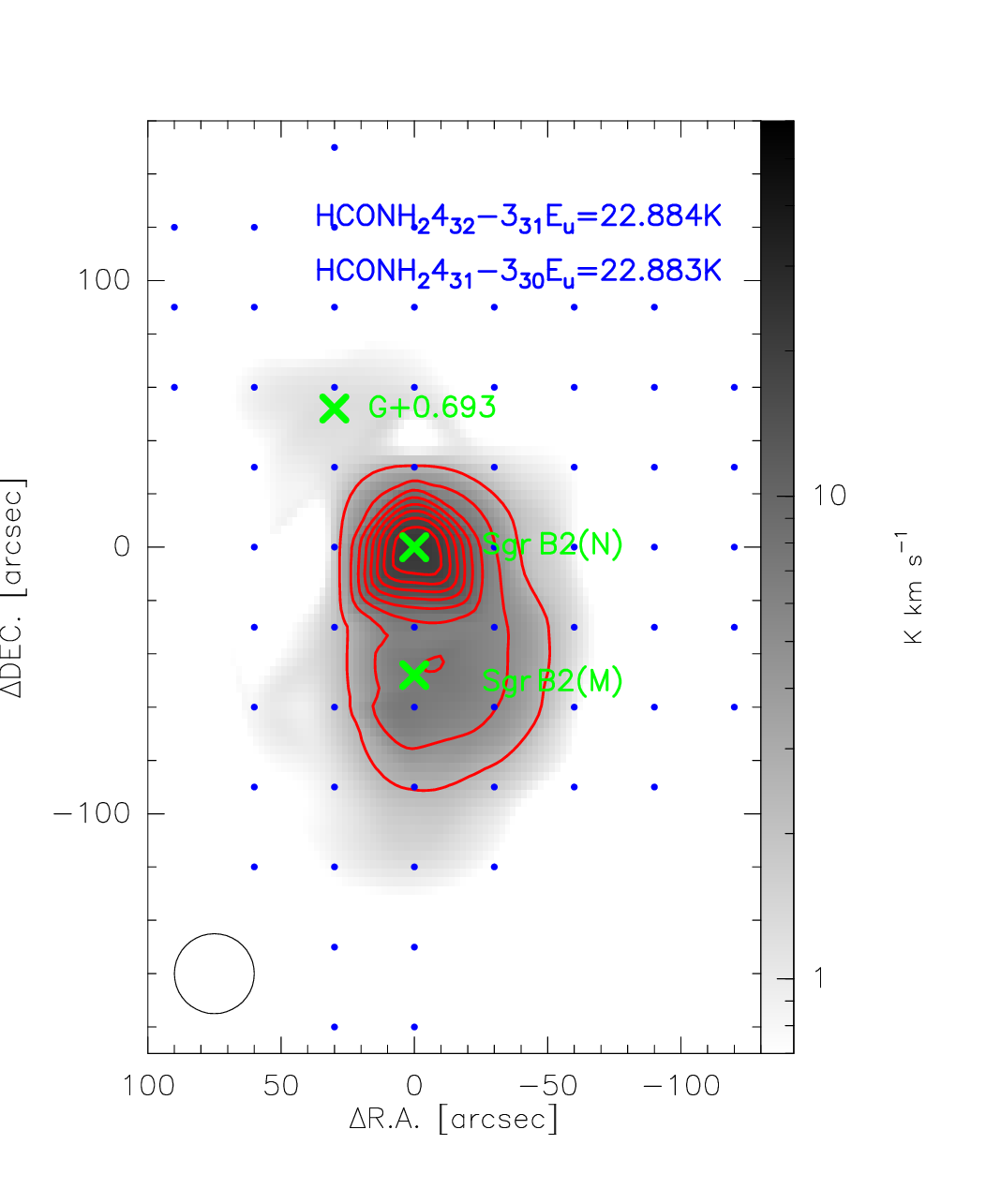}
    \caption{The gray-scale and contours show the spatial distribution of different transitions of HCONH$_2$. The figures are listed in order of increasing upper state energy. 
    The first figure is the map of HCONH$_2$ 1$_{1,1}$-0$_{0,0}$ at 82549.562 MHz. The contour levels start at 2 K km s$^{-1}$ with a step of 0.4 K km s$^{-1}$. 
    The second figure is the map of HCONH$_2$ 4$_{0,4}$-3$_{0,3}$ at 84542.330 MHz. The contour levels start at 7 K km s$^{-1}$ with a step of 2.5 K km s$^{-1}$. 
    The third figure is the map of HCONH$_2$ 4$_{1,3}$-3$_{1,2}$ at 87848.874 MHz. The contour levels start at 8.4 K km s$^{-1}$ with a step of 3 K km s$^{-1}$.
    The fourth figure is the map of HCONH$_2$ 4$_{2,2}$-3$_{2,1}$ at 85093.272 MHz. The contour levels start at 3.3 K km s$^{-1}$ with a step of 3 K km s$^{-1}$.
    The fiveth figure is the map of HCONH$_2$ 4$_{2,3}$-3$_{2,2}$ at 84807.795 MHz. The contour levels start at 3 K km s$^{-1}$ with a step of 2 K km s$^{-1}$.
    The sixth figure is the map of HCONH$_2$ 4$_{3,2}$-3$_{3,1}$ and 4$_{3,1}$-3$_{3,0}$ at 84888.994 MHz and 84890.987, respectively. The contour levels start at 2.5 K km s$^{-1}$ with a step of 2.5 K km s$^{-1}$.
 Sgr B2(N), Sgr B2(M), and G+0.693 are labeled with $''\times"$. The beam size are shown in the left corner of each panel.}
    \label{map_hconh2}
\end{figure}

The emission of HCONH$_2$ 1$_{1,1}$-0$_{0,0}$, 4$_{0,4}$-3$_{0,3}$, 4$_{2,3}$-3$_{2,2}$, 4$_{2,2}$-3$_{2,1}$ and 4$_{1,3}$-3$_{1,2}$ are observed to be stronger than 3 $\sigma$ in 60, 63, 59, 60 and 63 positions with 1 $\sigma$ of about 0.01 K in T$_{\rm mb}$, respectively. 
The emission of HCONH$_2$ 4$_{3,2}$-3$_{3,1}$ and 4$_{3,1}$-3$_{3,0}$ are observed to be weaker than other observed transitions of HCONH$_2$, which have the emission stronger than 3 $\sigma$ in 32 positions. 
The frequencies of these two transitions are very similar, with a difference of 2 MHz. As a result, these two transitions blended with each other, and  will be solved together. 
The detection threshold for all transitions is 3 sigma.
HCONH$_2$ 1$_{1,1}$-0$_{0,0}$ is partially blended with C$_2$H$_5$CN,v=1, but can be separated from each other in the spectra. This is also the only one b-type transition ($\Delta K_a = \Delta K_c = \pm1$), while others are a-type transitions ($\Delta K_a = 0$).
Since the profiles of the spectra in Sgr B2 are complex \citep[e.g.][]{2013A&A...559A..47B}, which can not be fitted with a single gaussian component, the integrated intensities over a certain velocity range would be used in the following works. 
The integrated intensities, 
the first moment (the intensity-weighted velocity) and the second moment (the intensity-weighted velocity dispersion) of the transitions in different positions are listed in Table \ref{intensity_hconh2_1} and Table \ref{intensity_hconh2_2}. 
The integrated intensity maps of each transition are shown in Figure \ref{map_hconh2} in order of increasing upper state energies. The intensities where the emission is lower than 3 $\sigma$ (1 $\sigma \sim$ 0.07 km s$^{-1}$ K) are assumed to be 0. Sgr B2(N) (including Sgr B2(N1), Sgr B2(N2) and so on), Sgr B2(M) and G+0.693, where HCONH$_2$ has been detected, are labeled with $''\times"$. Because the sampling does not satisfied the Nyquist sampling criterion, the emission sizes and peak positions for different transitions are decided by eyes, which are listed in Table \ref{emission_size}. The equivalent radius of the area where the emission is stronger than half of the emission at the peak position is used to describe the emission size.

\begin{figure}
    \centering
    \includegraphics[width=0.95\textwidth]{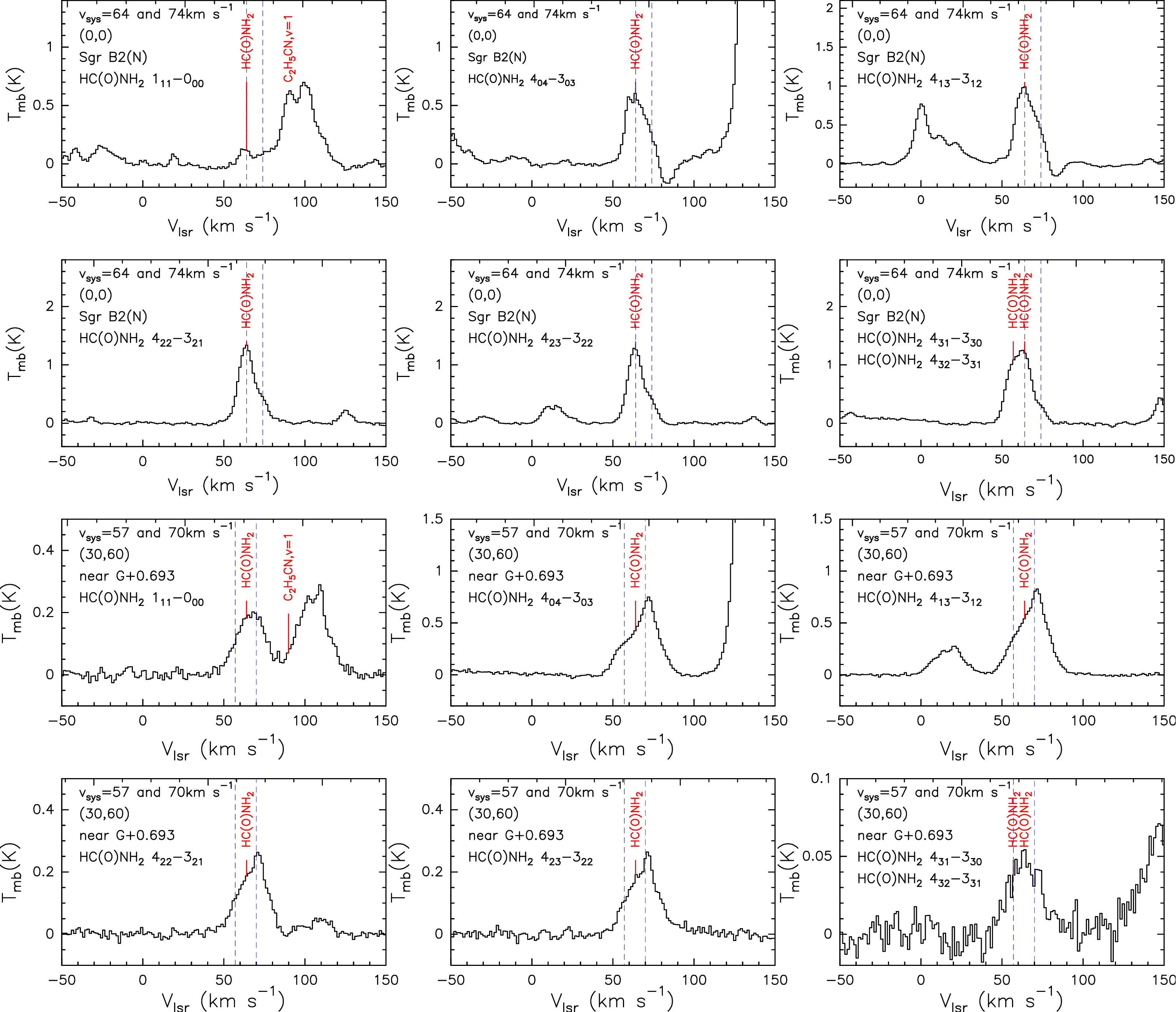}
    \caption{The spectra of HCONH$_2$ in position (30, 60) and Sgr B2(N). Frequencies are indicated with the red lines above the spectra. The $v_{\rm sys}$, determined by H$^{13}$CCCN 10-9, are labeled with blue dashed lines. The value of $v_{\rm sys}$ is labeled in the top left corner of each panel.}
    \label{spectra_hconh2}
\end{figure}

According to Table \ref{emission_size} and  Figure \ref{map_hconh2}, the emissions of different transitions are extended in Sgr B2, while the emission size decreases with the increasing upper state energies. 
The shift in peak position of different transitions is  significant comparing with the angular resolution of observations. The spectra of the observed transitions of HCONH$_2$ near the peak positions: Sgr B2(N) and position (30\arcsec, 60\arcsec) are displayed in Figure \ref{spectra_hconh2}, with blue dashed lines to show the LSR velocity. The LSR velocity in each position is determined with the line profile of H$^{13}$CCCN 10-9 in our data, considering its relatively strong emission and low optical depth.
For transition with a small upper state energy, corresponding to HCONH$_2$ 1$_{1,1}$-0$_{0,0}$, the emission has the largest emission size, and peaks at the north cloud of Sgr B2, near position (30\arcsec, 60\arcsec). G+0.693 is located in projection 0.32 pc (or 8\arcsec) south of the position (30\arcsec, 60\arcsec), with a LSR velocity of 68 km s$^{-1}$ \citep[][]{2006A&A...455..971R} and a typical lines FWHM of about 20 km s$^{-1}$ \citep[][]{2018MNRAS.478.2962Z}. The observed spectra in position (30\arcsec, 60\arcsec) (see Figure \ref{spectra_hconh2}) have a good agreement with this LSR velocity and linewidth. 
It has been suggested that there are large-scale, low velocity shocks \citep{2016MNRAS.457.2675H, 2020MNRAS.497.4896Z}. Many molecules have been found to be enhanced there, such as HNCO, HCO$_2^+$ and CH$_3$OH \citep[e.g.][]{1998ApJ...498..261M,2008MNRAS.386..117J}.
The low velocity shocks in G+0.693 might be a possible explanation for the emission peak there.
When the upper state energy increases, such as HCONH$_2$ 4$_{0,4}$-3$_{0,3}$ and 4$_{1,3}$-3$_{1,2}$, the peak moves to the west of Sgr B2(M), near position (-30\arcsec, -60\arcsec). The equivalent radius of emission region is $\sim$ 55\arcsec  for these two transitions. If the upper state energy is higher, like HCONH$_2$ 4$_{2,3}$-3$_{2,2}$, 4$_{2,2}$-3$_{2,1}$, HCONH$_2$ 4$_{3,2}$-3$_{3,1}$ and 4$_{3,1}$-3$_{3,0}$, the emission has the smaller emission size than other transitions, and peaks near Sgr B2(N). Sgr B2(N) is a source with many complex organic molecules (COMs) detected to have high column density and high excitation temperture, Most of the molecules there emit in the LSR velocity of $\sim$ 63-64 km s$^{-1}$ and 73-74  km s$^{-1}$, with a linewidth of 7 km s$^{-1}$ for each velocity component\citep[][]{2013A&A...559A..47B}. This is consistent with the spectra in Figure \ref{spectra_hconh2}.  Similar variation between the distribuion of different transitions has been mentioned by \cite{2008MNRAS.386..117J} and \cite{2011MNRAS.411.2293J}.
Such variation among different transitions reflects the complex excitation environment there, which would be analysed by rotation diagrams in Section \ref{subsubsec:Rd_diagram}.

\subsubsection{isotopic species}
\begin{figure}
    \centering
    \includegraphics[width=0.95\textwidth]{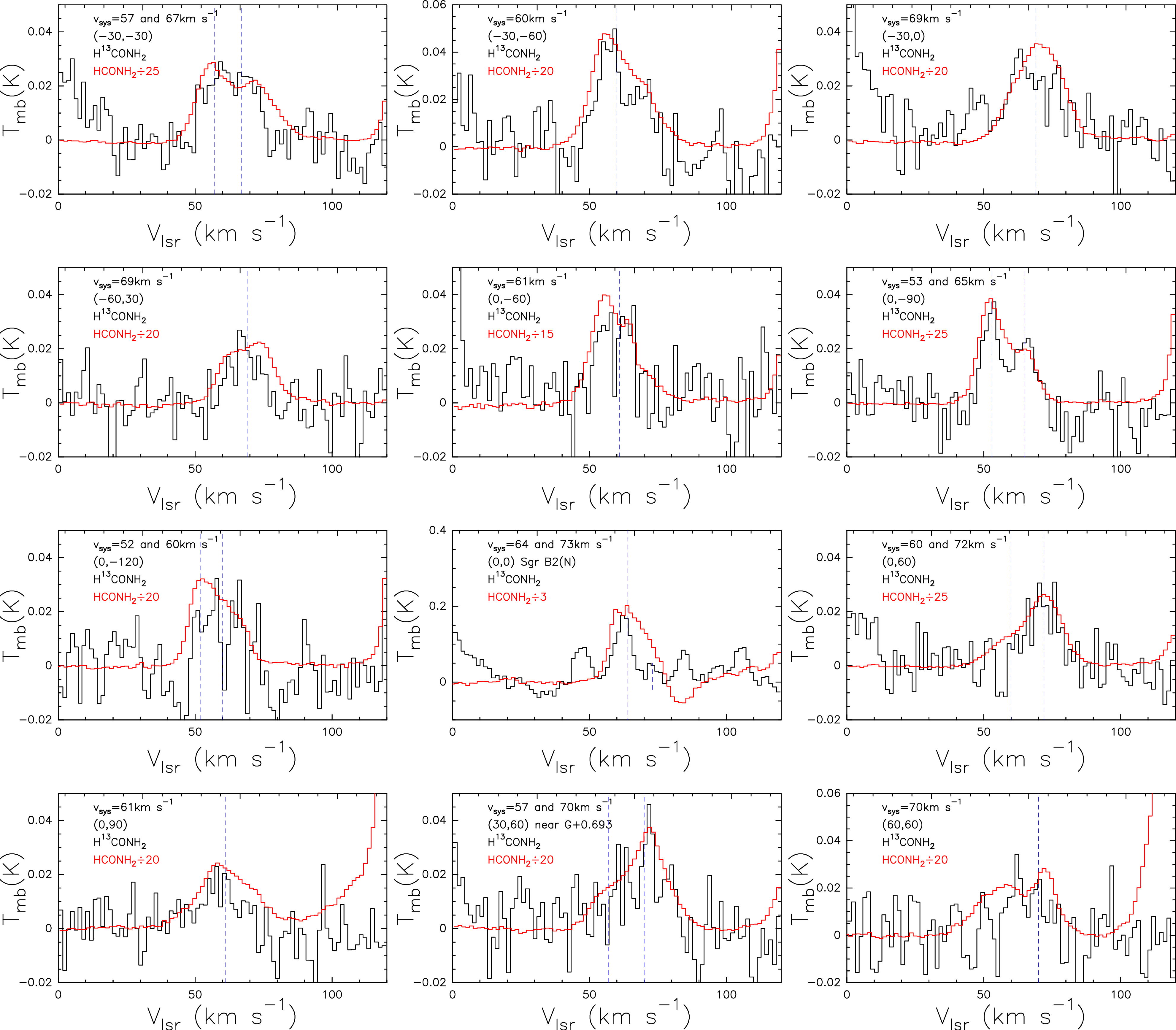}
    \caption{The spectra of HCONH$_2$ 4$_{04}$-3$_{03}$ and H$^{13}$CONH$_2$ 4$_{04}$-3$_{03}$. The $v_{\rm sys}$ are labeled with blue dashed lines. The spectra of HCONH$_2$ 4$_{04}$-3$_{03}$ are divided by 25, 20, 15 and 3 to compare with the spectra of H$^{13}$CONH2 in different positions.The value of $v_{\rm sys}$ is labeled in the top left corner of each panel.}
    \label{spec_13C}
\end{figure}

\begin{figure}
    \centering
    \includegraphics[width=0.5\textwidth]{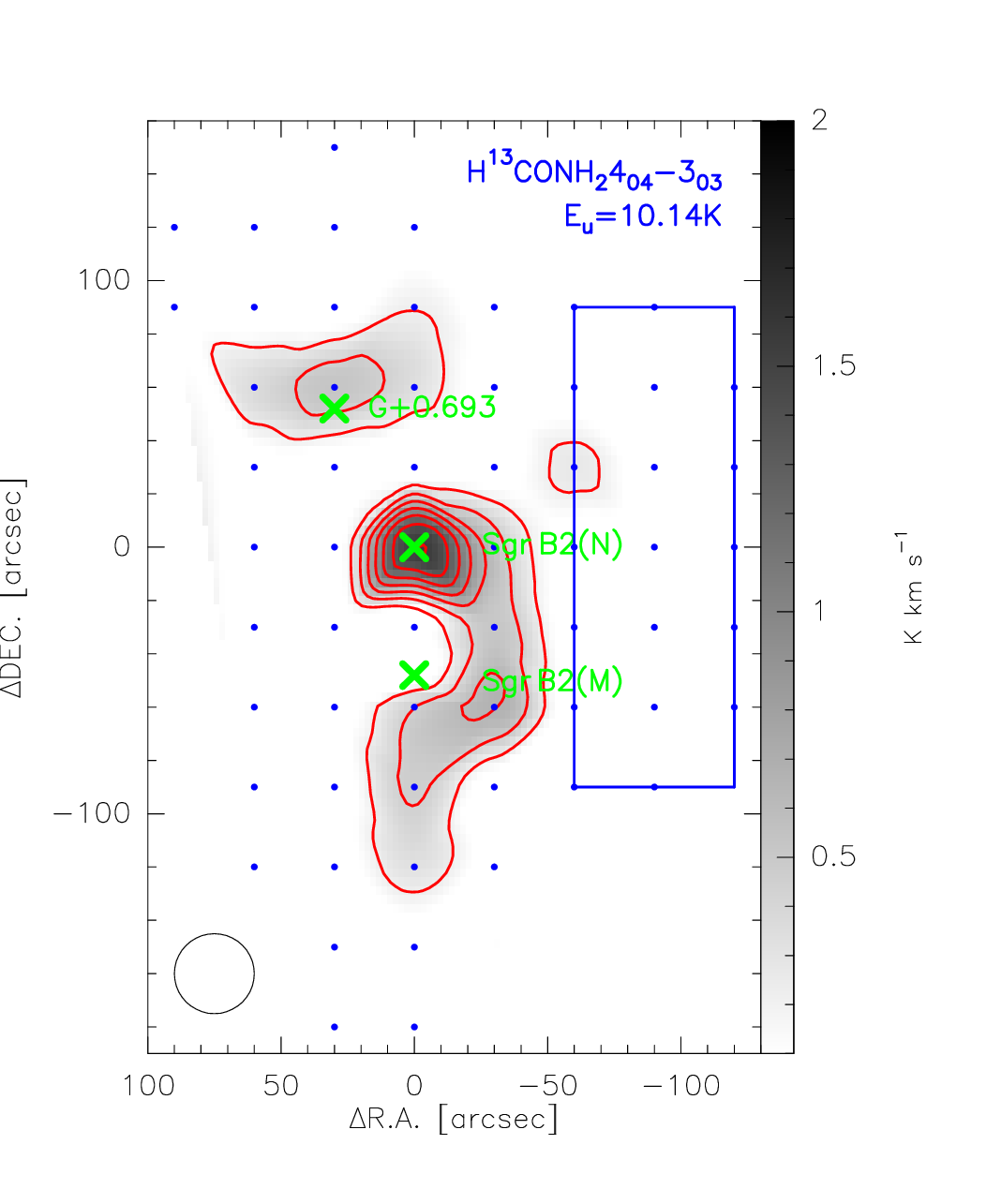}
    \caption{The distribution of H$^{13}$CONH$_2$ 4$_{04}$-3$_{03}$ at 84390.679 MHz. The contour map starts from 5 $\sigma$ with the step of 5 $\sigma$. 1 $\sigma$ is about 0.43 K km s$^{-1}$. Sgr B2(N) and Sgr B2(M) are labeled with $''\times"$. Beam size is shown in the bottom left corner. The area in the blue square shows the region used to average the spectra of the H$^{13}$CONH$_2$.}
    \label{distribution_13C}
\end{figure}

The emission of H$^{13}$CONH$_2$ 4$_{04}$-3$_{03}$ is also detected in the observation as listed in Table \ref{transition}.  In addition to Sgr B2(N), H$^{13}$CONH$_2$ 4$_{04}$-3$_{03}$ has an emission stronger than 3 $\sigma$ in 11 positions. The spectra are shown in Figure \ref{spec_13C}, with the spectra of HCONH$_2$ 4$_{04}$-3$_{03}$ as a comparison to investigate whether this transition is optically thick.
The spectra of HCONH$_2$ 4$_{04}$-3$_{03}$ are divided by 25, 20, 15 and 3 to compare with the spectra of H$^{13}$CONH$_2$ in different positions.
The $^{12}$C/$^{13}$C ratio can also be obtained from those two transitions. The integrated intensity map is drawn to find out its spatial distribution (see Figure \ref{distribution_13C}). The distribution of H$^{13}$CONH$_2$ 4$_{04}$-3$_{03}$ in the extended envelope is found to have a similar morphology with that of HCONH$_2$ 4$_{04}$-3$_{03}$. In comparison, the emission of H$^{13}$CONH$_2$ 4$_{04}$-3$_{03}$ peaks in Sgr B2(N) while that of HCONH$_2$ 4$_{04}$-3$_{03}$ is weak, impling the emission around there is optically thick. 

With the emission of isotope spieces detected in multiple positions, 
the isotope ratio  $^{12}$C/$^{13}$C in the region can be obtained from the intensity ratio of H$^{13}$CONH$_2$ and HCONH$_2$ 4$_{04}$-3$_{03}$.  
Here we assume that the abundance ratio of HCONH$_2$ to H$^{13}$CONH$_2$ has the same values as the intensity ratio, considering their similar upper state energies ($E_u$). 
Another assumption is that the ratio of $^{12}$C to $^{13}$C is equal to the abundance ratio of HCONH$_2$ and H$^{13}$CONH$_2$, namely, there is no isotope fraction. 
The isotope ratio is assumed to be a constant value in Sgr B2.
Based on these assumptions, the intensity ratio between HCONH$_2$ and H$^{13}$CONH$_2$ is equal to the ratio of $^{12}$C to $^{13}$C if the transitions used are optically thin. In Table \ref{13C_and_12C}, the intensity ratios of HCONH$_2$ 4$_{04}$-3$_{03}$ to H$^{13}$CONH$_2$ 4$_{04}$-3$_{03}$ in the positions where H$^{13}$CONH$_2$ 4$_{04}$-3$_{03}$ is detected are listed. Since HCONH$_2$ 4$_{04}$-3$_{03}$ is probably opically thick, a larger intensity ratio and the intensity ratio in the positions with large distance from hot cores can reflect the isotope ratio better. Therefore, position (0\arcsec, -60\arcsec) and (0\arcsec, 0\arcsec) are excluded when analysing the isotope ratio. Position (0\arcsec, 90\arcsec) is also excluded because of its extremely large intensity ratio.

\begin{figure}
    \centering
    \includegraphics[width=0.5\textwidth]{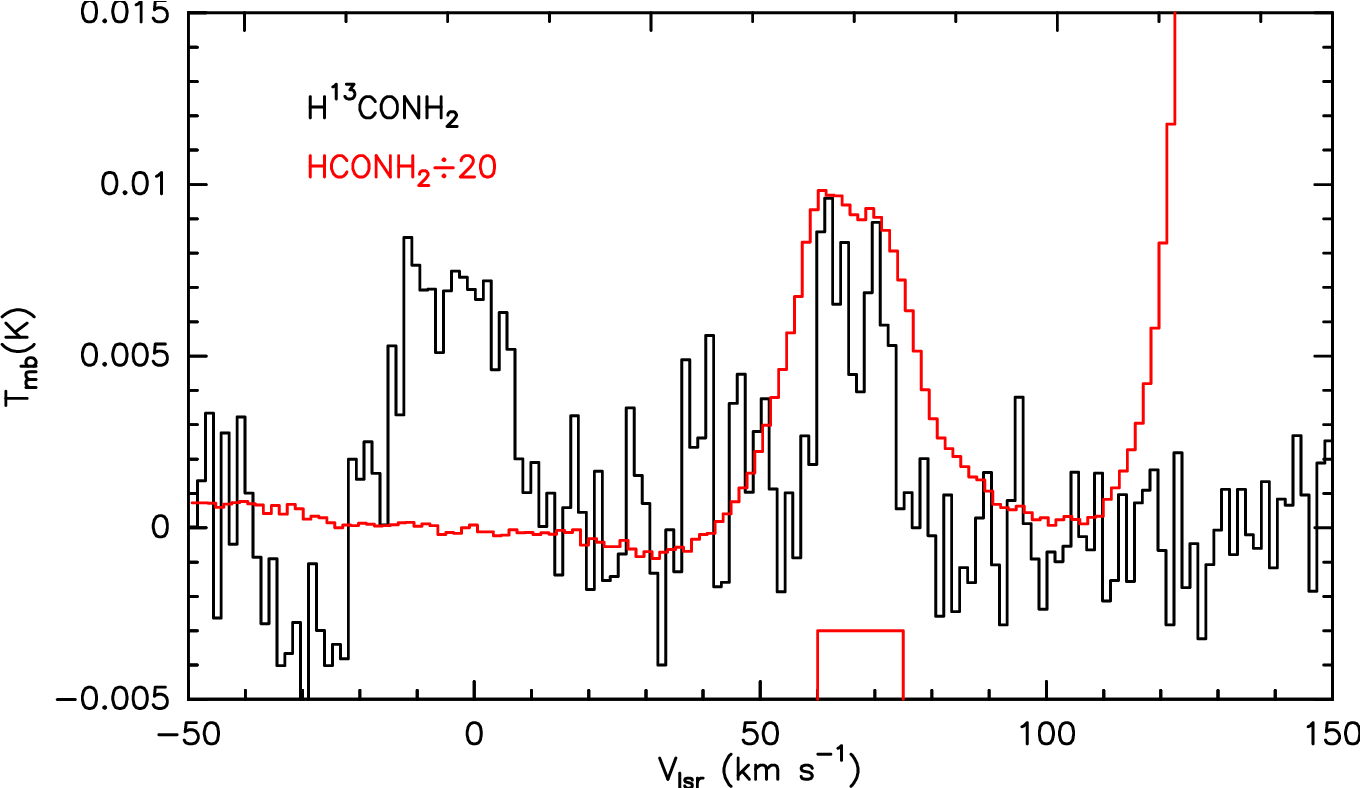}
    \caption{The averaged spectra of the H$^{13}$CONH$_2$ in the right region of Sgr B2. The red window represents the integrated velocity range, which is 60 km s$^{-1}$ to 75 km s$^{-1}$.}
    \label{spec_average}
\end{figure}

The ratio of $^{12}$C/$^{13}$C obtained from the intensity ratio is about 26 to 33, indicating the assumption of a constant ratio in the envelope of Sgr B2 is reasonable. To obtain the $^{12}$C/$^{13}$C ratio of the envelope, the emission of H$^{13}$CONH$_2$ 4$_{04}$-3$_{03}$ in the region far away from the hot cores are averaged. After averaging the spectra, the signal-to-noise ratio raises and the emission of H$^{13}$CONH$_2$ can than be detected to be stronger than 3 $\sigma$ (1 $\sigma$ = 1.8 mK) in the region shown within the blue square frame in Figure \ref{distribution_13C}. The averaged spectra of HCONH$_2$ 4$_{04}$-3$_{03}$ and H$^{13}$CONH$_2$ 4$_{04}$-3$_{03}$ are shown in Figure \ref{spec_average}. The spectra are integrated in the same velocity range which is marked with red window in the figure. The ratio between the intensities of these two lines is 28.7 $\pm$ 1.7 (listed in the last line of Table \ref{13C_and_12C}), within the uncertainty of the result calculated from single position, which can be regarded as the $^{12}$C/$^{13}$C ratio of the envelope.

The opical depths ($\tau$) of HCONH$_2$ 4$_{04}$-3$_{03}$ can then be calculated from the isotope ratio according to the following formula: 
\begin{equation}
    \frac{I_\mathrm{HCONH_2}}{I_\mathrm{H^{13}CONH_2}} = \frac{1-exp(-\tau)}{1-exp(\frac{\tau}{^{12}C/^{13}C})}
    \label{tau_formula}
\end{equation}. 
where $I_\mathrm{HCONH_2}$ and $I_\mathrm{H^{13}CONH_2}$ are the integrated intensities of HCONH$_2$ 4$_{04}$-3$_{03}$ and H$^{13}$CONH$_2$ 4$_{04}$-3$_{03}$. The results are listed in the last column of Table \ref{13C_and_12C}. 
In the envelope of Sgr B2, only the emission in position (0\arcsec, -60\arcsec) has a relatively large $\tau$ of 0.51. For the other positions, the emission is opically thin.
The much smaller ratio in position (0\arcsec, 0\arcsec), where Sgr B2(N) is located, gives a large $\tau$ of 5.30. Part of the emission there comes from the hot cores, which is difficult to be separated from the component of the envelope around. The strong continuum and high molecular abundance of the hot core may be a reason accounting for such larger $\tau$. Due to its complexity, we would not put emphasis on this position and may pay more attention to the positions in the cold extended envelope (further discussion in Section \ref{subsec:isotope}).

\subsubsection{Rotation diagram}\label{subsubsec:Rd_diagram}
As seven transitions of HCONH$_2$ have been detected in most of the regions in Sgr B2, the column densities and excitation temperature of HCONH$_2$ toward different positions can be attained with rotation diagrams.
The intensities of transition HCONH$_2$ 4$_{0,4}$-3$_{0,3}$ are corrected with the optical depth obtained from the $^{12}$C/$^{13}$C.
The other transitions are regarded to be opically thin under the LTE assumption, and a single temperature could be used to describe the energy level populations in the area. 

The transitions that are lower than 3 $\sigma$ levels would not be used in fitting, but they would be labeled in the diagrams with an arrow to represent their 3 $\sigma$ upper limits.  Rotation diagrams can be drawn according to the formula \citep{2009ARA&A..47..427H}:

 \begin{equation}
     \frac{N_u}{g_u} = \frac{N_{tot}}{Q(T_{rot})}e^{-\frac{E_u}{T_{rot}}}=\frac{8\pi k\upsilon ^2 I}{h c^3 A_{ul}g_u}
     \label{diagram_formula}
 \end{equation}
where N$_u$ is the column density of upper energy level, g$_u$ is the upper state degeneracy, N$_{tot}$ is the total column density, T$_{rot}$ is the rotation temperature, Q(T$_{rot}$) is the partition function, E$_u$ is the upper state energy, I is the intensity, c is the velocity of light, and A$_{ul}$ is the emission coefficient.
T$_{rot}$ is equal to excitation temperature (T$_{ex}$) under the LTE assumption. 
Q(T$_{rot}$) is the function of T$_{rot}$, calculated by the formula: $Q(T_{rot})=\sum g_i e^{E_i/kT_{rot}}$. The calibration error is not included into the error of intensities because the calibration error is small enough to be obmitted after averaging the spectra of each scan. Since the emission is more extended than the telescope beam according to the distribution map of different transitions, a filling factor correction is not considered.

The transition HCONH$_2$ 1$_{1,1}$-0$_{0,0}$ is not included in all of the diagrams because it deviated too much from the fitted straight line in the rotation diagram of every position.
This deviation can be the result of its different transition type from other lines, which is a b-type transition. It is likely that the a-type transitions and b-type transitions of HCONH$_2$ is not thermalized there. 
If the point of the b-type transition is located near the fitting line, the intensity of it in each position needs to be about 10 times weaker that what it is now. Therefore, the column density obtained from the b-type transition would be one magnitude larger than that obtained from a-type transitions. This phenomenon has also been observated by \cite{2000ApJS..128..213N} and \cite{2018MNRAS.478.2962Z}. 
Because the transitions of a-type and b-type couple together, namely both type of transitions can be produced from the same upper energy level, we regard the a-type and b-type as two different temperature components of HCONH$_2$. 
For position (30\arcsec, 60\arcsec), the excitation temperature of b-type transitions in G+0.693, which is 3.7 $\pm$ 0.7 K\citep[][]{2018MNRAS.478.2962Z}, can be used to calculate the column density of b-type transition there. The obtained column density of b-type transition there is 1.14 $\pm$ 0.23 $\times 10^{15} cm^{-2}$. The column density of a-type transitions derived from the rotation diagram is 1.3 $\pm$ 0.3 $\times 10^{14} cm^{-2}$ with an excitation temperature of 13.1 $\pm$ 3.0 K. The total column density in this position is 1.27 $\pm$ 0.26 $\times 10^{15} cm^{-2}$ taking into account both a-type and b-type transitions. The ratio of b-type to a-type transitons is 8.8. 
For other positions, only one b-type transition is detected. The rotation diagram is not available to deal with both types of the transition separately. So, we only give the column densities of HCONH$_2$ with a-type transitions. The significant difference betwween the excitation temperature and column density of both type transitions perhaps suggests that transitions with different types trace different physical regions.

In some positions, such as Sgr B2(N) and Sgr B2(M), there is absorption and seriously blending from other molecules. Therefore, these positions are not used to draw the rotation diagram. In addition, position (90\arcsec, 120\arcsec) and (90\arcsec, 90\arcsec) are also not used as these two positions would produce fake information in the distribution map.  
Position (30\arcsec, 150\arcsec), (-120\arcsec, -30\arcsec) and (-120\arcsec, -60\arcsec) are not used as well due to the low intensities of more than 2 transitions.

\begin{figure}
    \centering
    \includegraphics[width=0.4\textwidth]{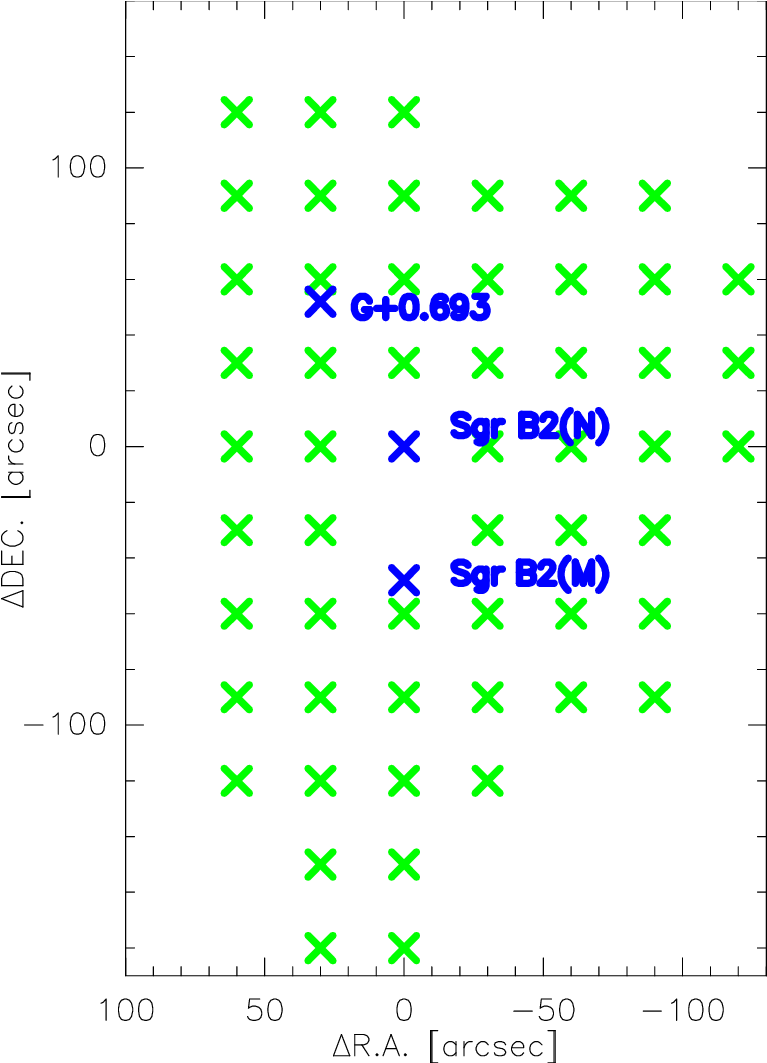}
    \caption{The positions selected to draw rotation diagrams. The selected position are labeled with green $''\times"$. Sgr B2(N), Sgr B2(M), and G+0.693 are labeled as blue $''\times"$.}
    \label{position}
\end{figure}

\begin{figure}
    \centering
    \includegraphics[width=0.45\textwidth]{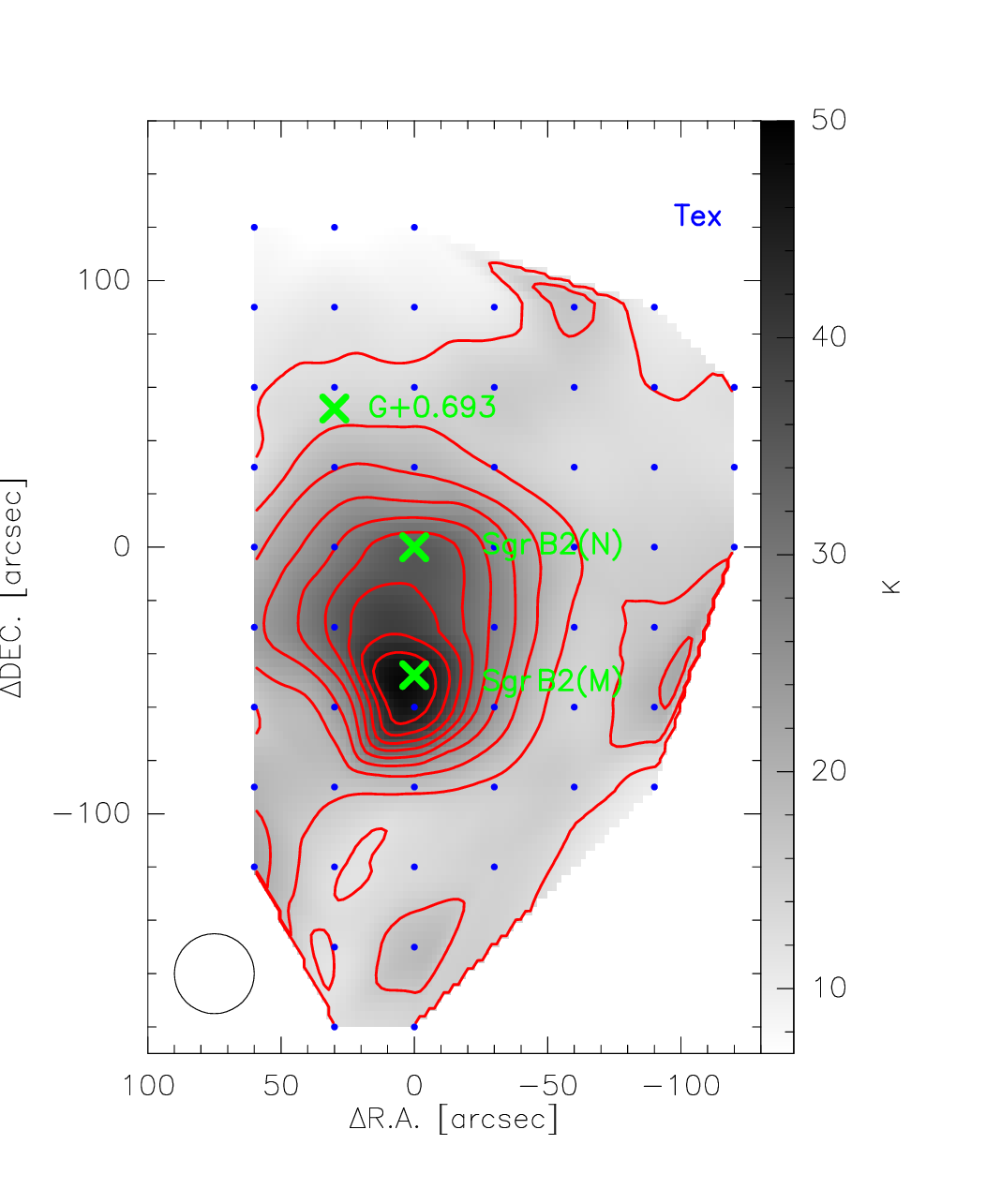}
    \includegraphics[width=0.45\textwidth]{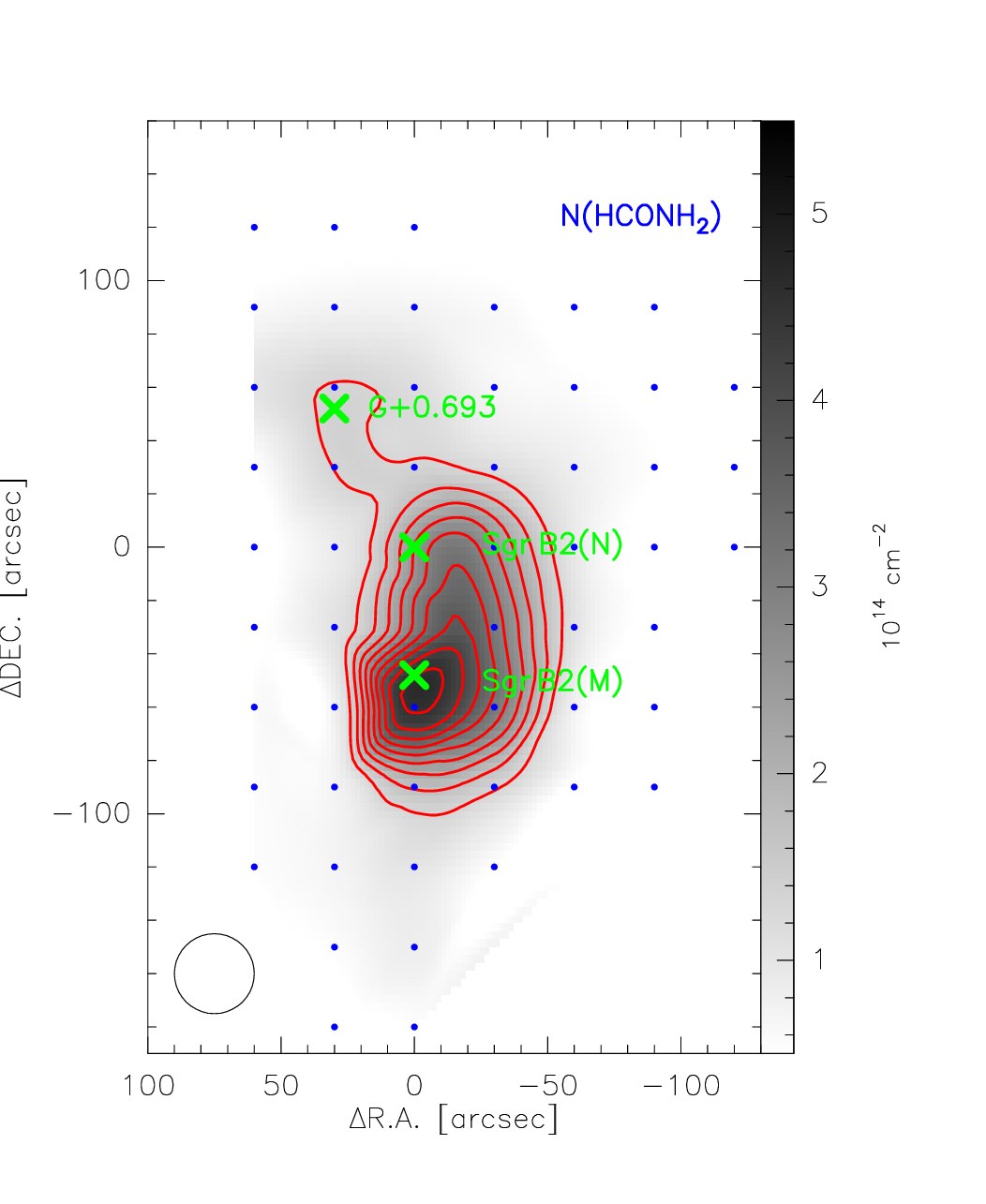}
    \caption{The distribution of excitation temperature and column density of HCONH$_2$. The contour levels of the left panel start at 12 K with a step of 4 K. The contour levels of the right panel start at 1.3 $\times 10^{14}$ cm$^{-2}$ with a step of 0.4 $\times 10^{13}$ cm$^{-2}$.}
    \label{tex}
\end{figure}

Only 54 positions are used to draw the rotation diagrams. The selected positions are shown with green $''\times"$ in Figure \ref{position}. The rotation diagrams are shown in Figure  \ref{ratotation_diagram} and Figure \ref{ratotation_diagram_3sigma}, corresponding to the positions with HCONH$_2$ 4$_3$-3$_2$ detected and not detected. Since HCONH$_2$ 1$_{1,1}$-0$_{0,0}$ is not included in the fit, it is labeled as an orange square, while other transitions are labeled as blue circles. 
From the results of ratotation diagrams, the distribution of column densities and excitation temperature could be obtained. The results are shown in Figure \ref{tex} and Table \ref{columndensity}. The values at the positions not labeled in Figure \ref{position} are interpolated from the values around them, which do not have physical meanings.
According to the figure, the $\rm T_{ex}$ of HCONH$_2$ range from 6 K to 46 K, with higher values around the hot cores. The column densities of HCONH$_2$ calculated with only a-type transitions range from 0.2 to 4.6 $\times 10^{14} cm^{-2}$. 
More details about the rotation diagrams are discussed in section \ref{subsec:rotation diagram}.

\subsection{CH$_3$CONH$_2$\label{subsec:ch3conh2}}

\subsubsection{distribution}

\begin{figure}
    \centering
    \includegraphics[width=0.95\textwidth]{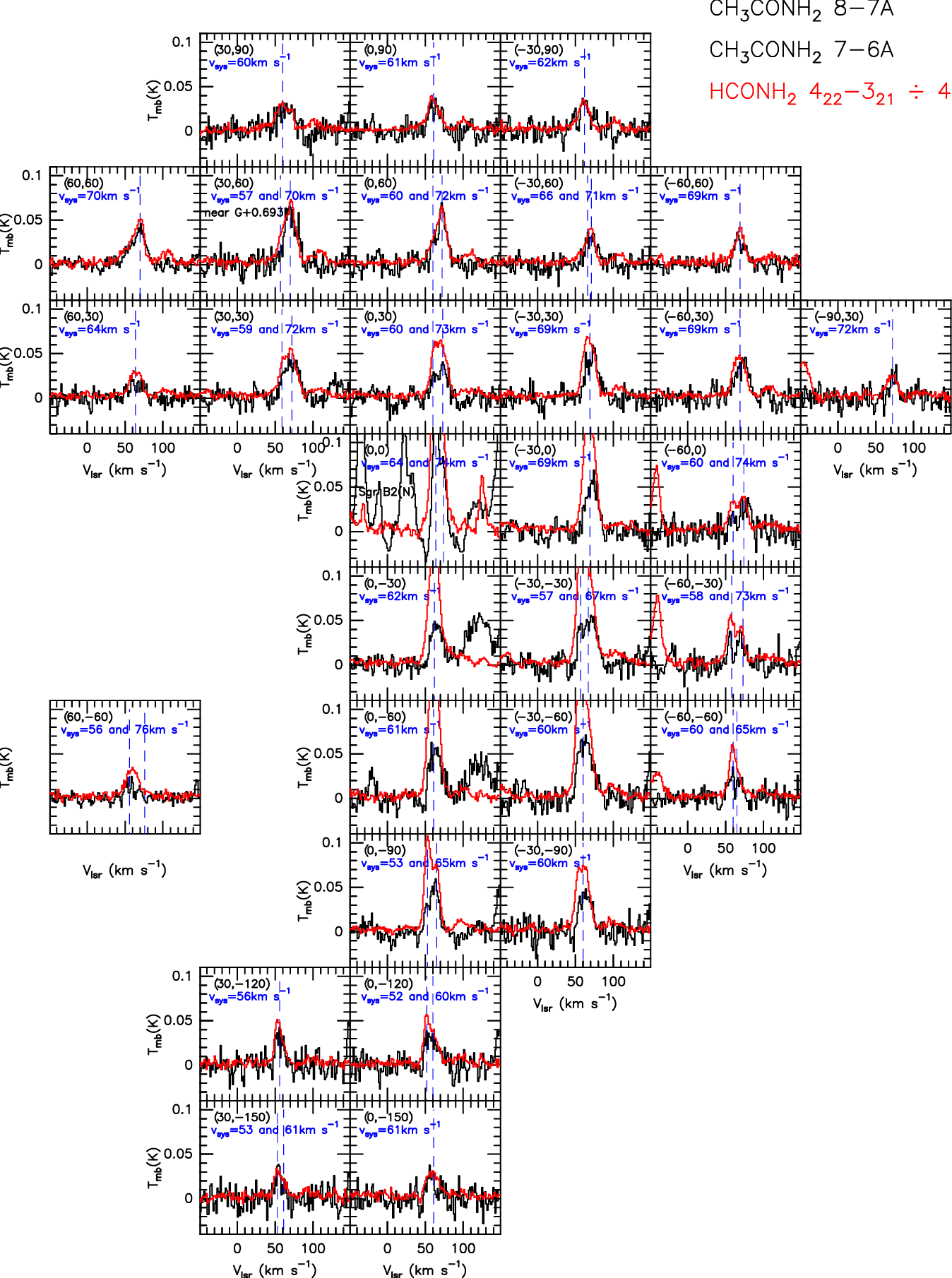}   
    \caption{The spectra of CH$_3$CONH$_2$ and HCONH$_2$ 4$_{22}$-3$_{21}$ in Sgr B2. The red profiles correspond to the spectra of HCONH$_2$ 4$_{22}$-3$_{21}$, while the black profiles are the spectra of CH$_3$CONH$_2$. The positions are labeled in the top left corner. The $v_{\rm sys}$ in difference positions are labeled with blue dashed lines. The value of $v_{\rm sys}$ is labeled  in the top left corner of each panel.}
    \label{spectra_ch3conh2}
\end{figure}

Because of the internal rotation of methyl group, CH$_3$CONH$_2$ has A- and E- species.  In our observations, only CH$_3$CONH$_2$-A has relatively clean and strong emission. The frequencies of the used CH$_3$CONH$_2$-A transitions are similar, with a difference of 2.7 MHz ($\sim$ 9.3 $km$ $s^{-1}$, a smaller line width than the typical line width in Sgr B2). Therefore, those transitions will also be used together. The column densities for all positions are measured with the emission of CH$_3$CONH$_2$-A. 
The transitions of CH$_3$CONH$_2$-A are observed to be stronger than 3 $\sigma$ in 30 positions. The spectra of CH$_2$CONH$_2$ in all the positions with emission stronger than 3 $\sigma$ levels are shown in Figure \ref{spectra_ch3conh2}, with the spectra of HCONH$_2$ 4$_{22}$-3$_{21}$ overlapped on them. The dashed blue lines are used to show the LSR velocity of the position. In the extended cold envelope, the profiles of the spectra of HCONH$_2$ and CH$_3$CONH$_2$ were similar to each other in the same position, from which we can assume these transitions were excited from the same cloud. 

\begin{figure}
    \centering
    \includegraphics[width=0.5\textwidth]{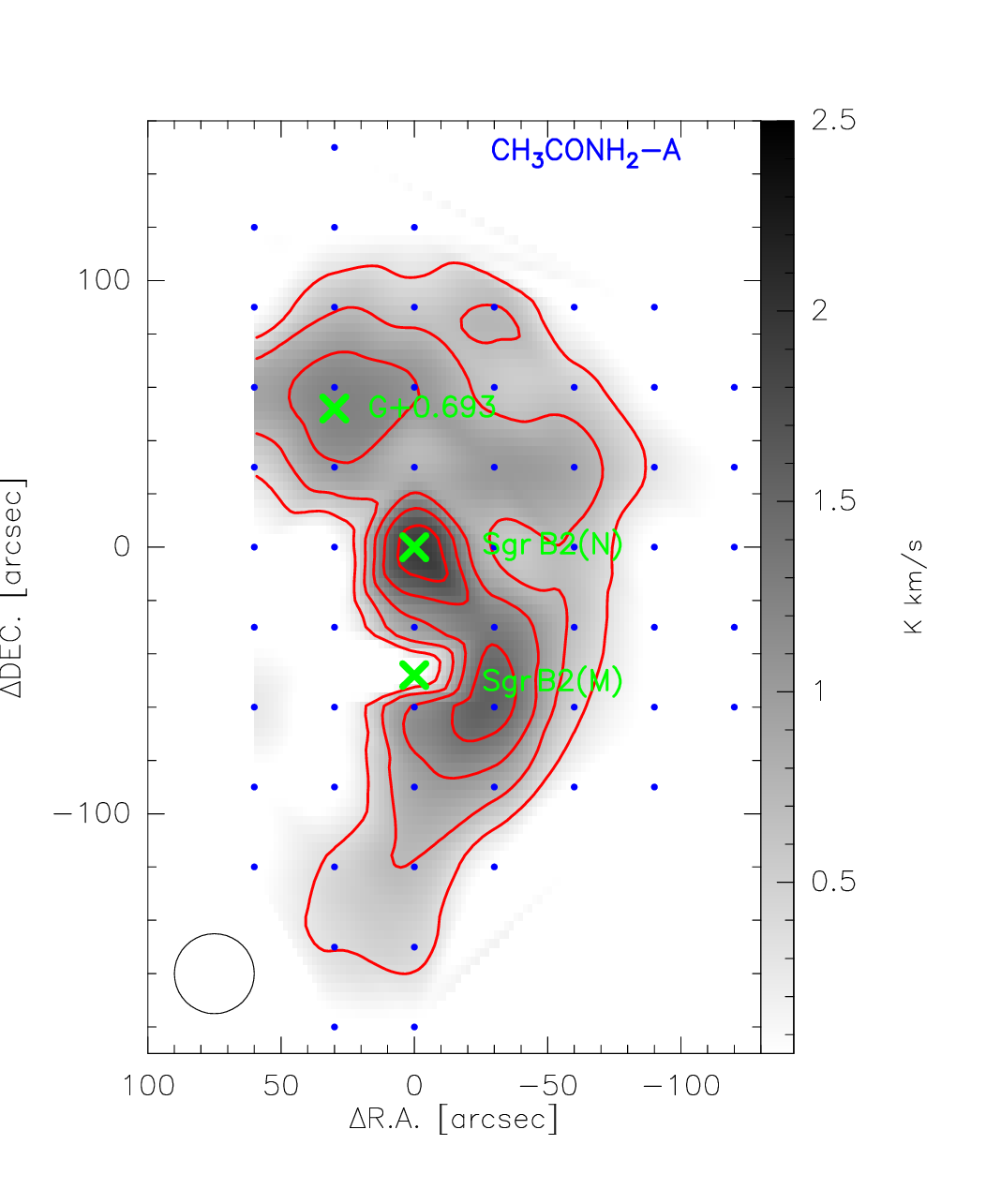}
    \caption{The integrated intensity map of CH$_3$CONH$_2$ emission. The integrated intensities of five transitions listed in Table \ref{transition} were used together to plot the map. The contour levels of CH$_3$CONH$_2$ emission start at 5 $\sigma$ with a step of 5 $\sigma$. $\sigma \approx$ 0.07 km s$^{-1}$ K for this map.  Sgr B2(N), Sgr B2(M), and G+0.693 are labeled with $''\times"$. Beam size is shown in the bottom left corner.}
    \label{map_ch3conh2}
\end{figure}

The integrated intensity map of CH$_3$CONH$_2$ is shown in Figure \ref{map_ch3conh2}.  All the transitions of CH$_3$CONH$_2$-A listed in Table \ref{transition} are used to draw the map. The intensities are assumed to be 0 in the positions where the emission is lower than 3 $\sigma$ levels. The emission size and the peak position are listed in Table \ref{emission_size}. The transitions of CH$_3$CONH$_2$ have similar emission sizes as that of HCONH$_2$ 4$_{0,4}$-3$_{0,3}$, while it peaks near Sgr B2(N). The distribution of emission in the envelope of Sgr B2 is similar to the transitions of HCONH$_2$, which may imply that these two molecules are chemically or physically linked. 

\subsubsection{Column density}

As the detected clean transitions of CH$_3$CONH$_2$ blend with each other, the column densities and excitation temperature can not be obtained by rotation diagrams. 
As a results, the column densities are calculated using the detected transitions together, assuming the same excitation temperature as HCONH$_2$. 

Since the difference between the upper state energies of different transitions is small, which is 19.82 K and 18.77 K for 8-7 A and 7-6 A respectively, they have a similar sensitivity to the excitation temperature. Namely, the column densities calculated from these transitions will not deviate too much if we assume they have the same upper state energy. Thus, a common upper state energy is used to simplify the calculation. Because the transitions CH$_3$CONH$_2$ 8-7 A have larger dipole moment (see in Table \ref{transition}), the emission of them would be stronger than that of CH$_3$CONH$_2$ 7-6 A. Most of the error is also generated by transitions CH$_3$CONH$_2$ 8-7 A. 
Therefore, the upper state energy of CH$_3$CONH$_2$ 8-7 A (19.82 K) is chosen as the upper state energy to avoid the complexity of calculation.  
The column densities are calculated with the following formula \citep{2007A&A...467..207P}: 
\begin{equation}
    N_{tot} = \frac{N_u}{g_u}\times Q(T_{ex})\times exp(E_u/kT_{ex}) 
    =\frac{3h}{8\pi^3S\mu^2} \times \frac{I\times Q(T_{ex}) }{J_{\nu}(T_{ex})-J_{\nu}(T_{bg})} \times \frac {J(T_{ex})}{\eta _{\nu}}
    \label{column_formula}
\end{equation}
where E$_u$ is the upper state energy, $\mu^2$ is the dipole moment, and S is the line strength. The value of $\mu^2S$ is obtained from the sum of each transition. $\eta _{\nu}$ is the beam filling factor, which is set to be 1 because the emission of $\rm CH_3CONH_2$ is extended.

\begin{figure}
    \centering
    \includegraphics[width=0.5\textwidth]{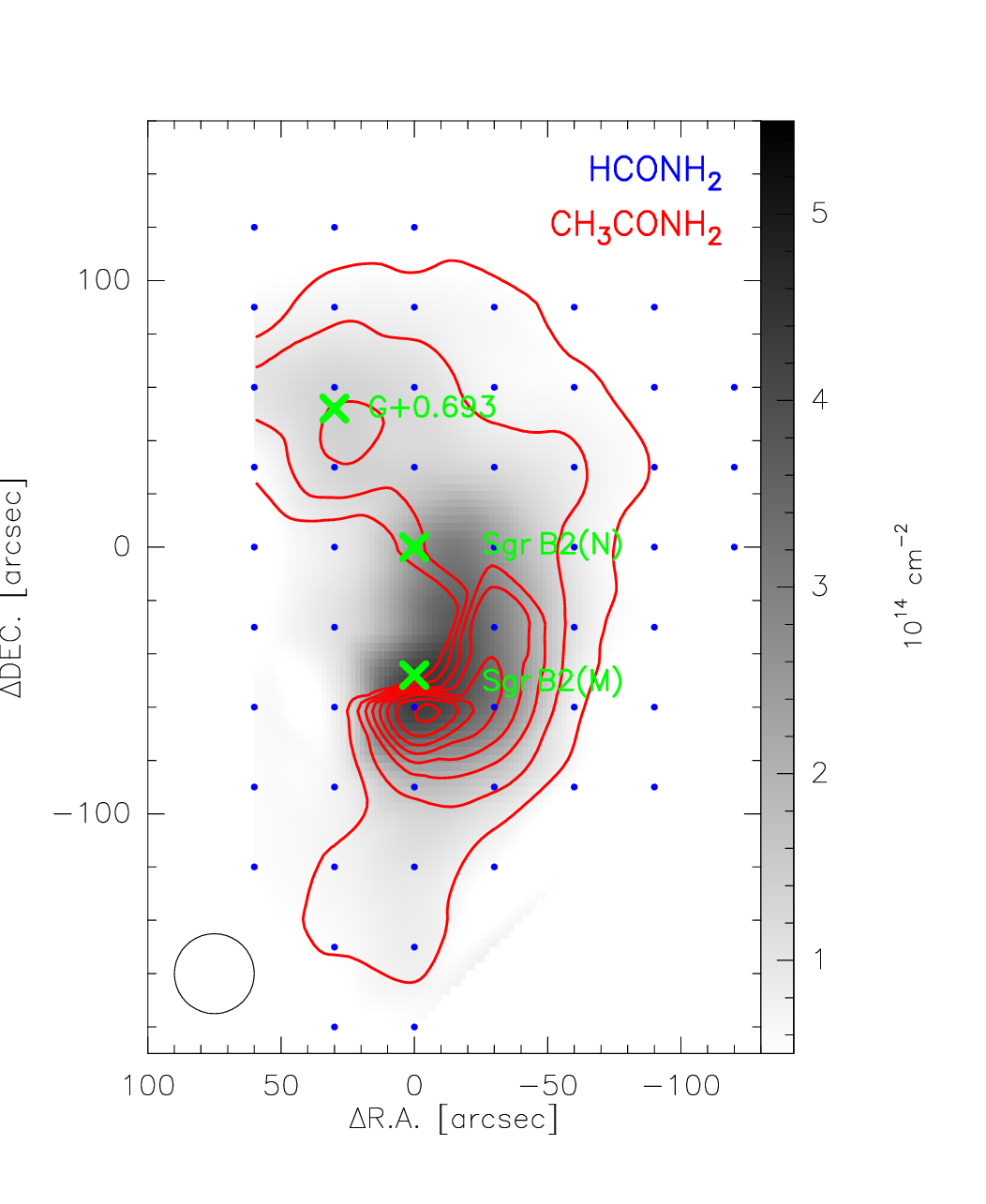}
    \caption{The distribution of column densities in Sgr B2. The gray-scale and the contour level represent the column density of HCONH$_2$ and CH$_3$CONH$_2$ respectively. The contour levels start at 0.7 $\times 10^{13}$ cm$^{-2}$ with the step of 1 $\times 10^{14}$ cm$^{-2}$.}
    \label{column_ch3}
\end{figure}

The results are listed in Table \ref{columndensity}. For the position where the emission of CH$_3$CONH$_2$ is not detected, the upper limits of column densities is given. 
The column densities of CH$_3$CONH$_2$ range from 0.6 to 7.6 $\times 10^{13} cm^{-2}$.
The distribution of column density is shown in Figure \ref{column_ch3}, with the distribution map of the column densities of HCONH$_2$ as comparison. The column densities where CH$_3$CONH$_2$ are not detected are set to be 0.  According to the figure, the distributions of both molecules have an extended distribution and increase rapidly around hot cores. The abundance ratio with respect to HCONH$_2$ is discussed in section \ref{subsec: ratio}.

\section{Discussion} \label{sec:discussion}

\subsection{isotopic species} \label{subsec:isotope}
From the intensity ratio of HCONH$_2$ to H$^{13}$CONH$_2$ (in Table \ref{13C_and_12C}), the derived ratio of $^{12}$C to $^{13}$C is 28.7 $\pm$ 1.7 in Sgr B2. 
This ratio is consistent with previous researches which obtain a ratio of 25 $\pm$ 8 and 31 $\pm$ 8 from the two components of HCONH$_2$ in Sgr B2(N) \citep{2017ApJ...845..158H}, and the average ratio of 24 $\pm$ 7 and 23 $\pm$ 6 obtained from the two components of 7 different molecules in Sgr B2(N), but slightly higher than the raito of $\sim$ 25 obtained from CH$_3$OH in Sgr B2 (N2) \citep{2016A&A...587A..92M}. 

However, this ratio is just the ratio obtained from the region in the west of Sgr B2. The emission of H$^{13}$CONH$_2$ 4$_{04}$-3$_{03}$ is too weak to be detected in the averaged spectra in other regions of Sgr B2. More observations are required to do a further study about the  $^{12}$C/$^{13}$C ratio in the envelope of Sgr B2.
\subsection{rotation diagram} \label{subsec:rotation diagram}
With the rotation diagrams of HCONH$_2$, the excitation temperature in the envelope of Sgr B2 is obtained, which ranges from 6 to 46 K. The rotation diagram of HCONH$_2$ has also been derived toward Sgr B2(N) by \cite{2011ApJ...743...60H}. They have detected a wide range of transitions with E$_u$ ranging from 10 K to 200K. Their results show that both molecules have a cold extended component and a hot component. In our observation, only transitions with low E$_u$ (\textless 40 K) are used to draw the rotation diagram. Therefore, only one cold component is obtained in each position.

Based on the results about the $^{12}$C and $^{13}$C in the section \ref{subsec:isotope}, the assumption that all the transitions of HCONH$_2$ are optically thin might not be correct in all positions.
If the emission is optically thick, the intensity of the transition would be underestimated, which would lead to an underestimated column density and an overestimated of $\rm T_{ex}$. For the positions near Sgr B2(N) and Sgr B2(M), an absorption led by the strong continuum there could also influence the results. 

In addition, drawing rotation diagrams need the assumption that $T_{bg}$ could be omitted comparing with $T_{ex}$ and the Rayleigh-Jeans law is valid (i.e. $h\nu \ll kT$), so that equation \ref{column_formula} could be transferred to equation \ref{diagram_formula}. However, $T_{ex}$ in positions at the boundary of our observing area, such as position (30\arcsec, 120\arcsec), the obtained excitation temperatures are lower than 10 K. Combined with the high frequencies of transitions, the Rayleigh-Jeans law can be hardly satisfied. In this case, the column densities obtained by the rotation diagrams would be underestimated.
The column densities were $\sim$ 30\%,$\sim$ 17\% and $\sim$ 10\% larger than the results obtained in section \ref{subsec:hconh2} when $T_{ex}$ is 7 K, 10 K, and 14 K, respectively. 

Besides, the rotation diagrams are drawn only with a-type transitions. The only detected b-type transition has a much stronger intensity than it should be if both types can reach thermalization. Thus, the excitation temperature of b-type transition can not be assumed to be same as that of a-type.  Here we only give the column densities based on the a-type transitions. To include b-type transitions into the analysis, more observations are needed.

It is found that the excitation temperatures in the envelope are smaller than the dust temperature, implying that these molecules are sub-thermalized. In this instance, a non-LTE method would be a better way to calculate the excitation temperatures and column densities.
However, the energy population can be well described with one single excitation temperature for the a-type transitions of HCONH$_2$. Therefore, the column densities are still reliable and non-LTE effect will not affect our results too much.

\subsection{Abundance ratio} \label{subsec: ratio}

To further study the relationship between CH$_3$CONH$_2$ and HCONH$_2$, the abundance ratio of them are calculated. HNCO would also be included in the discussion according to our former research \citep{2022RAA....22c5007Z}. Assuming the same excitation temperature as HCONH$_2$, the column densities could be obtained by the same way as CH$_3$CONH$_2$. In Sgr B2(M) and Sgr B2(N), there are many hot cores, as well as strong continuum emissions, which make the spectra difficult to be handled. Thus, the abundance ratio toward these two positions would not be included in the discussion.

\begin{figure}
    \centering
    \includegraphics[width=0.45\textwidth]{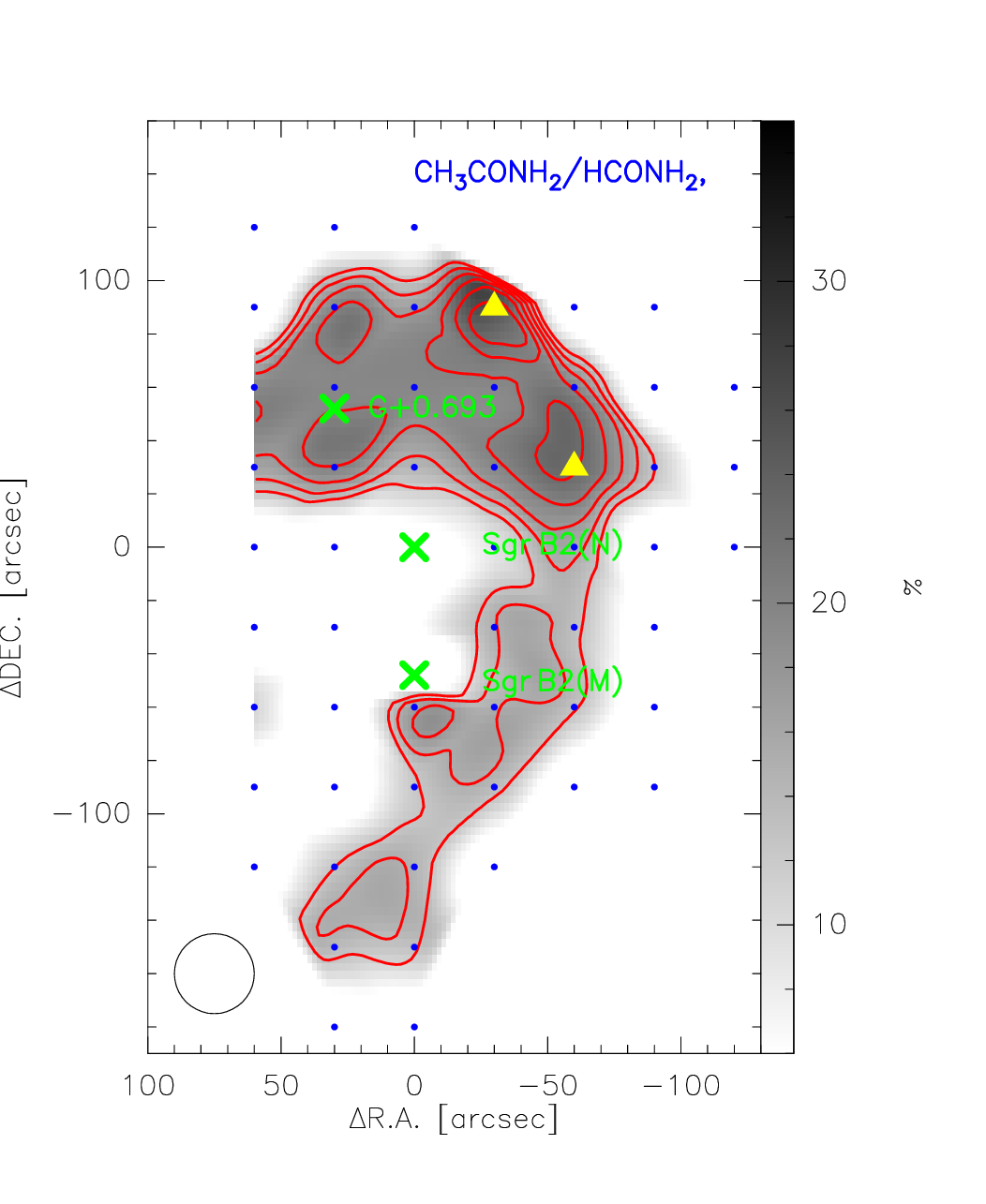}
    \includegraphics[width=0.45\textwidth]{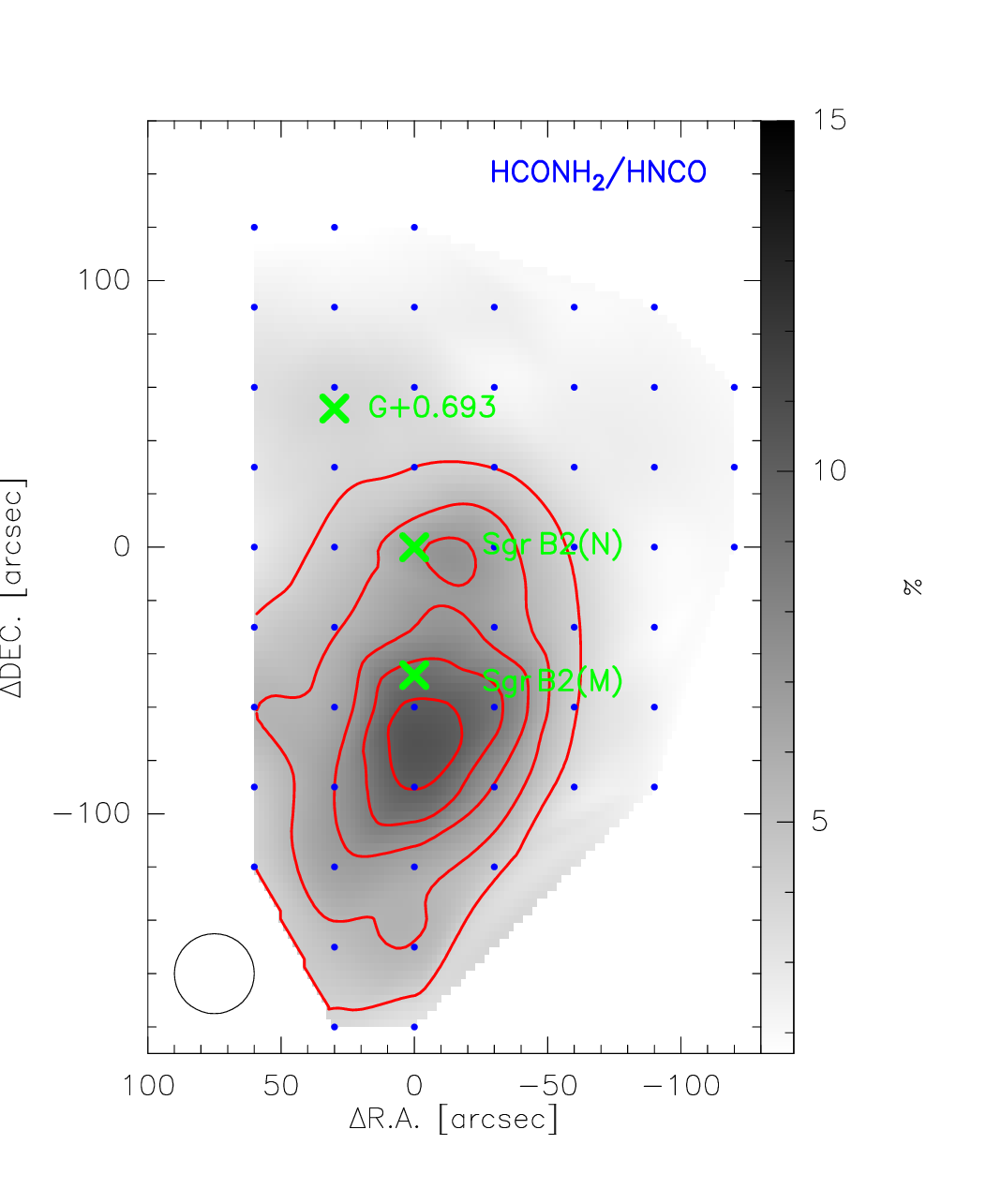}
    \caption{The distribution of the ratio between CH$_3$CONH$_2$, HCONH$_2$ and HNCO. The left figure corresponds to the ratio of CH$_3$CONH$_2$ to HCONH$_2$, of which the contour levels start at 12.5\% and increase in steps of 2.5\%. 
    The positions with the ratios larger than 20\% were pointed out with the yellow triangles.
    The figure in the right is ratio of HCONH$_2$ to HNCO, of which the contour levels start at 4\% and increase in steps of 1.5\%.}
    \label{distribution_ratio}
\end{figure}

The distribution maps of the column density ratio are displayed in Figure \ref{distribution_ratio}. 
From the figure, the abundance ratios between CH$_3$CONH$_2$ and HCONH$_2$ are found to range from 10 \% to 20 \%, except for one position with a smaller ratio, and two positions with larger ratios.
The two positions with the ratios larger than 20\% are marked with yellow triangles in the left panel of Figure \ref{distribution_ratio}. The spectra of those two molecules in these two positions are shown in Figure \ref{spectra_ch3conh2}. 
In those figures, the spectra of HCONH$_2$ 4$_{22}$-3$_{21}$ is chosen to compare with the spectra of CH$_3$CONH$_2$, because they have a similar E$_u$. Thus, the comparsion between those spectra can reflect the abundance ratio to some extent. 
The ratio of CH$_3$CONH$_2$ to HCONH$_2$ maximizes at position (-30\arcsec, 90\arcsec), with the values of 25.7\%. 
This constant ratio in most of the positions indicates that they have the same response to the physical or chemical environment. Perhaps, there is a chemical or physical link between CH$_3$CONH$_2$ and HCONH$_2$.
The ratio between HCONH$_2$ and HNCO range from 1.5\% to 10.5\%. 
The distribuion has a similar feature with the distribuion of excitation temperature, imply a possible relationship between the excitation temperature and the reactions that connect these two molecules.

\begin{figure}
    \centering
    \includegraphics[width=0.45\textwidth]{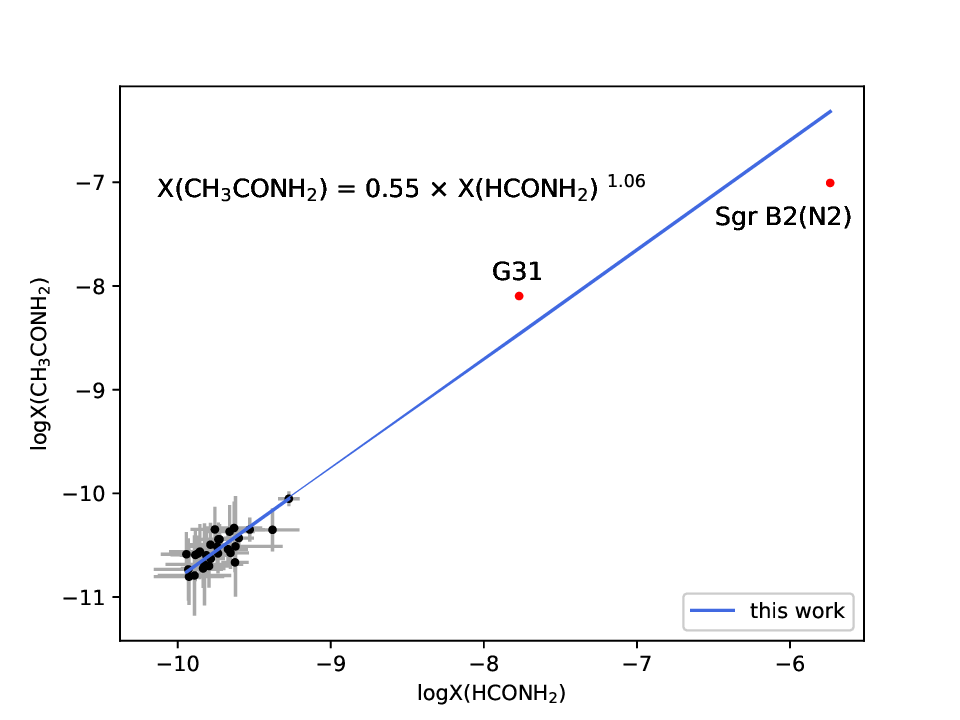}
    \includegraphics[width=0.45\textwidth]{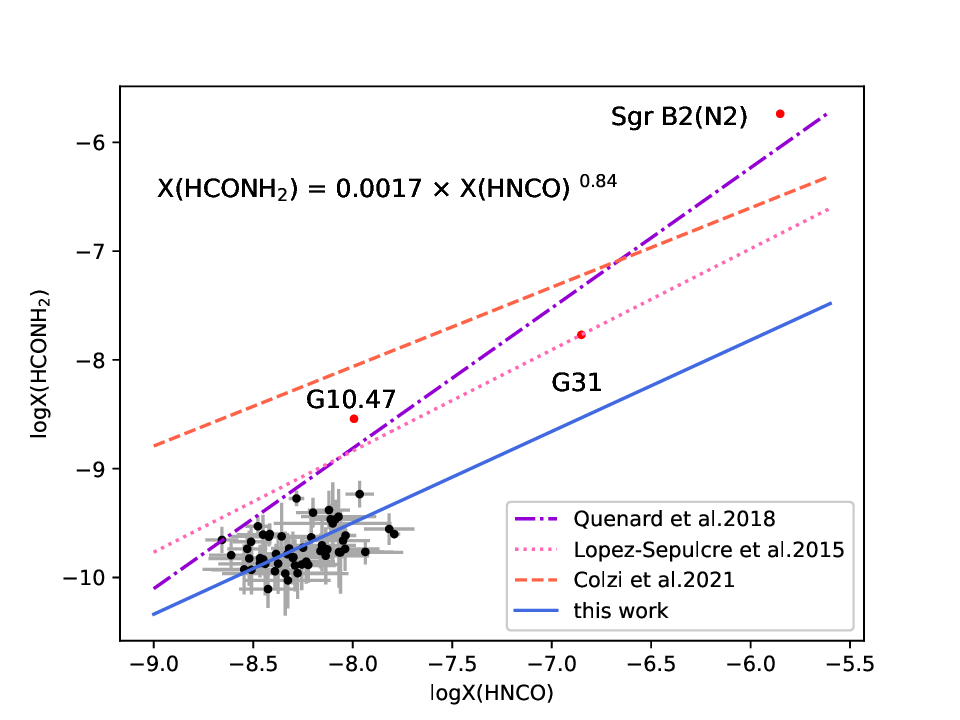}
    \caption{The abundance of HCONH$_2$, CH$_3$CONH$_2$, and HNCO, with respect to H$_2$. The blue lines are the power-law fitted results in our work. The dotted pink, dash-dotted purple and dashed orange line in the right panel are the results of \cite{2015MNRAS.449.2438L}, \cite{2018MNRAS.474.2796Q}, and \cite{2021A&A...653A.129C}. The red points in the figure show the abundance of HCONH$_2$ and HNCO in Sgr B2 (N2) \citep{2019A&A...628A..27B}, G10.47+0.03 \citep{2020ApJ...895...86G}, G31.41+0.31 \citep{2021A&A...653A.129C}. The black points are the observed abundances measured in the different positions of Sgr B2 complex. }
    \label{abundance_fit}
\end{figure}

To compare the results with other sources, the abundance with respect to H$_2$ are calculated. We use the Hi-GAL 350, 250, and 70 $\mu$m maps. the values are the averaged values in a 30\arcsec beam to match the beam of IRAM 30m at 3 mm. Figure \ref{abundance_fit} shows the abundances with respect to H$_2$. The blue lines are the power-law fits of the observed positions, which are given by the equation $X(\rm CH_3CONH_2)$=0.55 $\times$ $X(\rm HCONH_2)^{1.06}$ with a Pearson coefficient of 0.79, and $X(\rm HCONH_2)$=0.0017 $\times$ $X(\rm HNCO)^{0.84}$ with a Pearson coefficient of 0.39. The dotted pink, dash-dotted purple and dashed orange line correspond to the power-law fits given by \citet{2015MNRAS.449.2438L}, $X(\rm HCONH_2)$ = 0.04 $\times$ $X(\rm HNCO)^{0.93}$, by \citet{2018MNRAS.474.2796Q}, $X(\rm HCONH_2)$ = 32.14 $\times$ $X(\rm HNCO)^{1.29}$, and by \citet{2021A&A...653A.129C}, $X(\rm HCONH_2)$ = 0.006 $\times$ $X(\rm HNCO)^{0.73}$. The red points in the figure show the abundance of HCONH$_2$ and HNCO in Sgr B2 (N2) \citep{2019A&A...628A..27B}, G10.47+0.03 \citep{2020ApJ...895...86G}, G31.41+0.31 \citep{2021A&A...653A.129C}.
The data used to fit the pink dotted lines in \cite{2015MNRAS.449.2438L} include only the compact/inner solution of rotation diagram analyses in shocks and star formation regions with different masses. The sources used in \cite{2018MNRAS.474.2796Q} include shocks and hot corinos.
The line obtained in our work lies below the other three lines, implying a relatively low abundance of HCONH$_2$ comparing with HNCO in the extended envelope. The low Pearson coefficient of 0.39 also indicates the weak relationship between HNCO and HCONH$_2$. It is likely that the formation mechanism of these molecules are different in the envelope compared with the mechanism in the compact sources. 
This conclusion has a good agreement with the lower panel of Figure 2 of \cite{2015MNRAS.449.2438L}.
For the line between HCONH$_2$ and CH$_3$CONH$_2$, a nearly linear relationship needs to be explained with an updated chemical model.

Some uncertainties may affect the final results. First, the critical densities vary with molecules, leading to the difference in excitation temperature. If the excitation temperature of CH$_3$CONH$_2$ is lower than HCONH$_2$, it would result in a lower column density of CH$_3$CONH$_2$ and a smaller abundance ratio. Besides, only the clean transitions of CH$_3$CONH$_2$-A are detected. The emission of CH$_3$CONH$_2$-E, which may help to attain more accurate column density, is too weak or blend with other molecules. 
What's more, the low signal to noise ratio due to the low abundance of CH$_3$CONH$_2$ may also affect the abundance ratio obtained from the integrated intensity.  Therefore, an observation with higher resolution, higher sensitivity and  wider bandwith are needed in the future.

\subsection{Some special positions}
Sgr B2(N): It is a massive star-formation region with many compact molecular cores and strong cotinuum. Since there is strong continuum, the spectra in Sgr B2(N) are not analyzed. The results in Sgr B2(N) drawn in the figures that describe the spatial distribution of excitation temperature and abundance ratio are the results of interpolation based the position besides, and thus can not reflect the exact physical condition there. From the integrated intensity maps of the transitions of both molecules, the peaks for transitions with relatively higher upper state energies and the peak of the transition of isotopic species are located in Sgr B2(N), which indicates the high column density here.
\cite{2011ApJ...743...60H} found that the column densities of HCONH$_2$ and CH$_3$CONH$_2$ towards Sgr B2(N) are 5.6 $\times$ 10$^{14}$ cm$^{-2}$ and 6.9 $\times$ 10$^{14}$ cm$^{-2}$. Therefore, the column densities of HCONH$_2$ and CH$_3$CONH$_2$ should peak in Sgr B2(N).
For the ratio of HCONH$_2$/HNCO, the results attained by ALMA show that the column density of HNCO, HCONH$_2$, CH$_3$CONH$_2$ in Sgr B2(N2) is 2 $\times 10^{18}$ $cm^{-2}$, 2.6 $\times 10^{18}$ $cm^{-2}$, and 1.4 $\times 10^{17}$ $cm^{-2}$, respectively \citep{2017A&A...601A..49B}. The ratio of HCONH$_2$/HNCO is 130\%. In Sgr B2(N3), Sgr B2(N4), Sgr B2(N5), the abundance ratio of HCONH$_2$ to HNCO is 25\%, \textless 56\% and 44\%, respectively\citep{2017A&A...604A..60B}. 
The abundance ratio is higher than our results, implying a different mechanism there. The trend that the ratio of HCONH$_2$ to HNCO increases with the excitation temperature (see Figure \ref{tex} and Figure\ref{distribution_ratio}) is consistent with the larger ratio found in Sgr B2(N). In terms of CH$_3$CONH$_2$/HCONH$_2$, the abundance ratio in Sgr B2(N1E) and Sgr B2(N1) is 9\% and 15\%\citep{2021ApJ...919....4L}. In Sgr B2(N2), \cite{2017A&A...601A..49B} found the abundance ratio of CH$_3$CONH$_2$/HCONH$_2$ to be 5.4\%.
In Sgr B2(N1S), the abundance ratio of CH$_3$CONH$_2$/HCONH$_2$ is 14\% \citep{2019A&A...628A..10B}. The abundance ratio obtained from the hot core is consistent with that in the cold envelope.

Sgr B2(M): It is a hot core at a later stage than Sgr B2(N). The continuum there is stronger and consist of more dense cores than Sgr B2(N)\citep{2016A&A...588A.143S}.  The linewidths of the spectra there range from $\sim$ 6 to 15 km s$^{-1}$ and the LSR velocity is 60 to 66 km s$^{-1}$ \citep[][]{2013A&A...559A..47B}. An obvious absorption in the spectrum of HCONH$_2$ 4$_{04}$-3$_{03}$ and a slight absorption in the spectrum of HCONH$_2$ 4$_{13}$-3$_{12}$ are detected there, while CH$_3$CONH$_2$ was not detected. Perhaps the CH$_3$CONH$_2$ here has been destroyed or the spectrum have strong absorption.

G+0.693: Among our observation positions, (30\arcsec, 60\arcsec) is the nearest from G+0.693. In position (30\arcsec, 60\arcsec), the column density of CH$_3$CONH$_2$ is 2.6 $\times$ 10$^{13}$ cm$^{-2}$, while HCONH$_2$ is 1.3 $\times$ 10$^{14}$ cm$^{-2}$. The abundance ratio of HCONH$_2$/HNCO and CH$_3$CONH$_2$/HCONH$_2$ is 3.9\% and 19.2\%, respectively.
In the observation of \citet{2018MNRAS.478.2962Z}, the ratio of HCONH$_2$/HNCO is 20\%. If only a-HCONH$_2$ is considered, the ratio is 1.6\%.
In the work of \citet{2023MNRAS.523.1448Z}, the column density of HCONH$_2$ is calculated from the isotope species H$^{13}$CONH$_2$, which is 2.5 $\times$ 10$^{14}$ cm$^{-2}$. The $^{12}$C/$^{13}$C ratio they used is 40, which is larger than our result of 28.7. The column density of HCONH$_2$ is consequently larger than the column density we obtain. Because multiple transitions of CH$_3$CONH$_2$, they obtain the  excitation temperature of 7.4 K and 7.8 K for CH$_3$CONH$_2$-A and CH$_3$CONH$_2$-E, lower than the temperature we used. If the same excitation temperature is used, a similar column density of CH$_3$CONH$_2$-A can be obtained. The abundance ratio in their work is 46\%, considering both CH$_3$CONH$_2$-A and CH$_3$CONH$_2$-E.


\subsection{Chemical model} \label{subsec:model}

\begin{figure*}
        \begin{center}
           \includegraphics[width=1.0\textwidth]{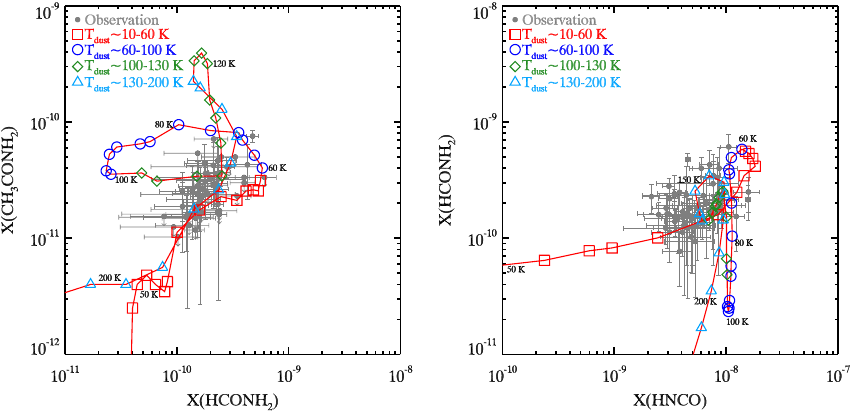}
           \end{center}
        \caption{Comparison of observed abundances of CH$_3$CONH$_2$, HCONH$_2$, and HNCO (with respect to H$_2$) with simulated abundances using the best-fit chemical model in \citet{2021A&A...648A..72W}. The red solid lines represent the evolution of $X$(CH$_3$CONH$_2$) versus $X$(HCONH$_2$) and $X$(HCONH$_2$) versus $X$(HNCO) at different time in the warm-up phase; the dust temperature at a few selected points are labelled. The results for $T_{\mathrm{dust}}\sim10-60 \, \mathrm{K}$,  $60-100 \, \mathrm{K}$,  $100-130 \, \mathrm{K}$, and  $130-200 \, \mathrm{K}$ are plotted as squares, circles, diamonds, and triangles, respectively. The best-fit model is an evolving model with the following parameters: $n_{\mathrm{H}}=2\times10^3 \, \mathrm{cm^{-3}}$, $A_{\mathrm{V}}=10$, the duration of the cold phase $t_{\mathrm{cold}}=3\times10^5 \, \mathrm{yr}$, the duration of the warm-up phase $t_{\mathrm{warmup}}=2\times10^5 \, \mathrm{yr}$, the maximum temperature after the warm-up $T_{\mathrm{max}}=200 \, \mathrm{K}$, the external UV radiation scaling factor $G_0=10^4$ with an X-ray burst at $T_{\mathrm{gas}}=T_{\mathrm{dust}}=20 \, \mathrm{K}$ (the total timescale $t_{\mathrm{total}}=3.459\times10^5 \, \mathrm{yr}$), the effective ionization rate $\zeta_{\mathrm{CR}}=1.3\times10^{-13} \, \mathrm{s^{-1}}$ and the duration $t=10^2 \, \mathrm{yr}$ (for details see Table 1 and Section 3.2.2 in \citet{2021A&A...648A..72W}). }
        \label{figure14}
        \end{figure*}

To investigate the synthesis mechanism of peptide-like molecules HCONH$_2$ and CH$_3$CONH$_2$ in the region extending among $100\arcsec$ from Sgr B2, we study the evolution of HCONH$_2$ and CH$_3$CONH$_2$ in chemical models. The simulations provided by \citet{2021A&A...648A..72W} are adopted for comparing with observations, since most observed positions still locate in the low-density envelope around Sgr B2 \citep{1993A&A...276..445H, 1995A&A...294..667H, 2016A&A...588A.143S}, for which the chemical evolution had been calculated via Monte Carlo method and explained different distributions of seven complex organic molecules (COMs). Besides CH$_2$OHCHO, CH$_3$OCHO, t-HCOOH, C$_2$H$_5$OH, CH$_3$NH$_2$, CH$_3$OCH$_3$, and C$_2$H$_5$CN analyzed by \citet{2021A&A...648A..72W}, HCONH$_2$ and CH$_3$CONH$_2$ are already included in the chemical reaction network. Figure \ref{figure14} presents the relationship between the abundance $X$(CH$_3$CONH$_2$) (with respect to H$_2$) and $X$(HCONH$_2$), and the relationship between $X$(HCONH$_2$) and $X$(HNCO) in the best-fit model (the same one as proposed by \citet{2021A&A...648A..72W}) comparing with observations, while the other models cannot fit observations. The corresponding physical evolution is that the low density extended region around Sgr B2 ($n_{\mathrm{H}}=2\times10^3 \, \mathrm{cm^{-3}}$) underwent a cold phase with gas and dust temperature $T_{\mathrm{gas}}=T_{\mathrm{dust}}=10 \, \mathrm{K}$, then a warm-up phase of increasing both temperatures simultaneously to reach the maximum 200 K, including an X-ray flare from Sgr A* with a short duration of no more than 100 yr when $T_{\mathrm{gas}}=T_{\mathrm{dust}}=20 \, \mathrm{K}$ (for details see Table 1 and Section 3.2.2 in \citet{2021A&A...648A..72W}). Under such circumstance, simulated $X$(CH$_3$CONH$_2$), $X$(HCONH$_2$), and $X$(HNCO) can match observed abundances at $T_{\mathrm{dust}}\sim53-56 \, \mathrm{K}$, $105-115$ K, and $170-190$ K, respectively. However, considering the relatively low dust temperature $T_{\mathrm{dust}}\sim20-30 \, \mathrm{K}$ in the extended envelope according to the dust continuum, our chemical models still need to be further modified.

We leave rectification of the chemical reaction network for a future detailed study. Here we only discuss those reactions that are key for the species concern us here and need to be further assessed. According to current simulations, the most important reactions for producing HCONH$_2$, CH$_3$CONH$_2$, and HNCO at $T_{\mathrm{dust}}\sim53-56 \, \mathrm{K}$ are JNH$_2$ + JHCO $\rightarrow$ HCONH$_2$, JNH$_2$ + JCH$_3$CO $\rightarrow$ CH$_3$CONH$_2$, and JHNCO $\rightarrow$ HNCO, respectively, and the letter J represents species on the grain. Therefore, the reactive desorption driven by exothermic reactions \citep{2007A&A...467.1103G, 2013ApJ...769...34V} is the most significant mechanism for producing HCONH$_2$ and CH$_3$CONH$_2$ in relatively cold environment, which is also concluded for the other seven COMs \citep{2021A&A...648A..72W}. For HNCO which has already been synthesized abundantly on the grain surface, the thermal desorption is the main mechanism for increasing its gaseous abundance since its evaporation temperature is about 50 K. 

Similarly, the thermal desorption induces rapidly increasing abundances of HCONH$_2$ and CH$_3$CONH$_2$ at $T_{\mathrm{dust}}\sim105-115 \, \mathrm{K}$, which exceeds their evaporation temperatures (about 93 K and 105 K, respectively, calculated with the binding energies of 5556 K and 6281 K presented by \citet{2013ApJ...765...60G} and via equation (6) presented by \citet{2012A&A...538A..91D}). In addition, the gaseous ion-molecule reactions involving C$^+$ are the main approach for depleting HCONH$_2$ and CH$_3$CONH$_2$ molecules in the warm-up phase, which is similar to seven COMs \citep{2021A&A...648A..72W}. These reactions may explain the evolution of HCONH$_2$ and CH$_3$CONH$_2$ in the region close to Sgr B2(N) and Sgr B2(M) with higher density and temperature. Nevertheless, for producing enough HCONH$_2$ and CH$_3$CONH$_2$ molecules at $T_{\mathrm{dust}}\sim20-30 \, \mathrm{K}$ in the extended envelope, more efficient non-thermal desorption mechanisms need to be investigated in the future analysis, not only reactive desorption but also photodesorption and Eley-Rideal mechanism \citep{2007A&A...467.1103G, 2010ApJ...716..825O, 2012ApJ...759L..43C, 2013ApJ...769...34V, 2014ApJ...795L...2V, 2015MNRAS.447.4004R}. 

Besides non-thermal desorption processes, other reasons can cause the discrepancy between best-fit dust temperatures derived from simulations and observations. 1) The uncertainties of many parameters for calculating reaction rates in chemical models are large, especially the binding energy (or desorption energy) of each species \citep{2017ApJ...844...71P}. 
For example, in this chemical reaction network, the adopted binding energies of HCONH$_2$ and HNCO are 5556 K and 2850 K originating from OSU network database and \citet{2013ApJ...765...60G}, while new calculated values 5468 K and 4684 K using the Gaussian 09 suite of programs show a smaller difference \citep{2020ApJ...895...86G}, which may induce their evolution to become similar. 
Additionly, contrast to the adopted fix values in most chemical models for simplicity, the binding energies measured by laboratory experiments vary with the coverage of grain surface \citep{2016ApJ...825...89H, 2019ApJ...875...73B}. The binding energy distributions are also inferred by theoretical computations \citep{2020ApJ...904...11F, 2022ApJ...938..158P} and significantly influence the efficiencies of thermal desorption and grain surface two-body reactions \citep{2017ApJ...844...71P, 2020A&A...643A.155G}, which may increase the abundances of HCONH$_2$ and CH$_3$CONH$_2$ at $T_{\mathrm{dust}}\sim20-30 \, \mathrm{K}$. 2) Some new reactions for synthesizing HCONH$_2$ and CH$_3$CONH$_2$ may have not been considered in aforementioned models. One reaction is the gas-phase reaction NH$_2$ + H$_2$CO $\rightarrow$ HCONH$_2$ + H, which has been used to explain the observation toward the shock region L1157-B1\citep{2017A&A...605L...3C}. The dual-cyclic hydrogen addition and abstraction reactions between HNCO and HCONH$_2$ in ice grains derived from experiments \citep{doi:10.1021/jacs.9b04491} are not included, neither. Thus, more accurate chemical simulations adopting binding energy distributions and new reactions are necessary to fit observed abundances in the envelope around Sgr B2 with low dust temperature in the future study. 3) The aforementioned models have not adopted multiple physical mechanisms simultaneously to investigate whether an individual mechanism can induce a good fit comparing with observations. However, the environment of Sgr B2 is complex, more sophisticated physical model including not only X-ray flare from Sgr A* \citep{2023Natur.619...41M} but also shocks should be considered in simulations. Such accurate physical model demands more observations to limit it.





\section{Summary} \label{sec:summary}

We have performed mapping observation toward Sgr B2 to further study the spatial distributions of the peptide-like molecules and the relationship between them. The obtained results are listed there:

1. HCONH$_2$ and CH$_3$CONH$_2$ are detected to have an extended distribution toward Sgr B2, with the abundance ratio range from 10\% to 20\%. The abundance of  CH$_3$CONH$_2$ was enhanced toward to north-west of Sgr B2. The excitation temperature in the envelop of Sgr B2 ranges from 6K to 46 K, based on the rotation diagram of HCONH$_2$.

2. The isotopic molecule H$^{13}$CONH$_2$ is also detected toward 12 positions of Sgr B2. The emission of H$^{13}$CONH$_2$ 4$_{04}$-3$_{03}$ indicated that HCONH$_2$ 4$_{04}$-3$_{03}$ has absorption toward Sgr B2(N). The ratio of $^{12}$C/$^{13}$C is 28.7 toward Sgr B2 envelope.

3. A chemical model including a cold phase and a warm-up phase with a short duration X-ray burst of no more than 100 yr at $T_{\mathrm{gas}}=T_{\mathrm{dust}}=20 \, \mathrm{K}$ is used to explain the observed abundances of HCONH$_2$ and CH$_3$CONH$_2$. The simulations can fit observed abundances at $T_{\mathrm{dust}}\sim53-56 \, \mathrm{K}$, $105-115$ K, and $170-190$ K, respectively. The reactive desorption driven by exothermic reactions dominate the synthesis of these two peptide-like molecules in relatively cold environment. Such dust temperatures are still higher than observations, which requires more accurate chemical models for explaining it.

\begin{acknowledgments}
    This work is based on observations carried out under project number 170-18 with the IRAM 30m telescope. IRAM is supported by INSU/CNRS (France), MPG (Germany) and IGN (Spain). This work has been supported by the National Key R\&D Program of China (No. 2022YFA1603101). This work is also supported by CDMS database. We thank JinJin Xie for the suggestions for the draft. YW acknowledges the support by the Natural Science Foundation of Jiangsu Province (Grant Number BK20221163). 
\end{acknowledgments}

\clearpage

\clearpage

\clearpage

\clearpage

\clearpage

\clearpage

\clearpage

\clearpage

\clearpage

\clearpage

\clearpage

\clearpage

\clearpage

\begin{table}
    \begin{center}
        \caption{The transitions of HCONH$_2$ and CH$_3$CONH$_2$}
        \label{transition}
        \begin{tabular}{lccccclc}
            \hline
            \hline
            Molecule &Transition & Rest Freq.& $E_u$ & log$_{10}$(A) & $\mu^2S$ & Comments  & Reference \\
                &       &  MHz  &   K   & s$^{-1}$  &  $D^2$ & & \\
            \hline
            HCONH$_2$ & 1$_{1,1}$-0$_{0,0}$ & 82549.562 & 3.962 &  -5.8001	  & 0.726 & Partially blend with C$_2$H$_5$CN,v=1. &   \\
                    & 4$_{0,4}$-3$_{0,3}$ & 84542.330 & 10.16 & -4.3888 & 52.279 & No blend.  &  \\  
                    & 4$_{2,3}$-3$_{2,2}$ & 84807.795 & 22.10 &  -4.5094  & 39.227 & No blend. &  \\
                    & 4$_{3,2}$-3$_{3,1}$ & 84888.994 & 37.00 &  -4.7422    & 22.884 & No blend. & (1) \\
                    & 4$_{3,1}$-3$_{3,0}$ & 84890.987 & 37.01 &  -4.7422    & 22.883 & No blend.&  \\
                    & 4$_{2,2}$-3$_{2,1}$ & 85093.272 & 22.12 &  -4.5051  & 39.227 & No blend.&  \\
                    & 4$_{1,3}$-3$_{1,2}$ & 87848.874 & 13.52 &   	-4.3666    & 49.030 & No blend. &  \\
            \hline
            H$^{13}$CONH$_2$ & 4$_{0,4}$-3$_{0,3}$ & 84390.679 & 10.14 & -4.3912 & 52.259 & No blend.&  (1)  \\
            \hline
            CH$_3$CONH$_2$-A & 7$_{2,6,0}$-6$_{1,5,0}$ & 87629.757 & 18.77 & - & 65.860 & No blend. &  \\
                              & 8$_{0,8,0}$-7$_{1,7,0}$ & 87632.434 & 19.82 & - &  94.010 & No blend.&   \\
                              & 8$_{1,8,0}$-7$_{1,7,0}$ & 87632.443 & 19.82 & - &  5.868 & No blend. &  (2)\\
                              & 8$_{0,8,0}$-7$_{0,7,0}$ & 87632.501 & 19.82 & - &  5.868 & No blend. &   \\
                              & 8$_{1,8,0}$-7$_{0,7,0}$ & 87632.510 & 19.82 & - &  94.010 & No blend. &  \\
            \hline
        \end{tabular}
    \end{center}
    Note: The transitions of HCONH$_2$ and CH$_3$CONH$_2$ used for mapping. For CH$_3$CONH$_2$, only A-spieces were detected. Col. (1) chemical formula; Col. (2) transition quantum number \citep{2013A&A...559A..47B}; Col. (3) rest frequency; Col. (4): upper state energy level (K); Col. (5): emission coefficient; Col. (6): $\mu^2$ is the dipole moment, and S is the line strength; Col. (7) comments.

    Reference: (1)\cite{2009JMoSp.254...28K}, (2) \cite{2004JMoSp.227..115I}
    

\end{table}

\clearpage
\begin{table}
    \begin{center}
        \caption{The emission size and peak position of HCONH$_2$ and CH$3$CONH$_2$}
        \label{emission_size}
        \begin{tabular}{cccc}
            \hline
            molecule & transition & peak position & emission size \\
                &       &              &    \arcsec  \\
            HCONH$_2$ & 1$_{1,1}$-0$_{0,0}$  & (19, 47) &    81  \\
               & 4$_{0,4}$-3$_{0,3}$  & (-29, -55) &   53   \\
               & 4$_{1,3}$-3$_{1,2}$  & (-29, -52) &   57   \\
                & 4$_{2,2}$-3$_{2,1}$  & (-4, -4) &  37  \\
                & 4$_{2,3}$-3$_{2,2}$  & (-5, -5) &   38  \\
                & 4$_{3,2}$-3$_{3,1}$  & (0, 0) &  20  \\
                & 4$_{3,1}$-3$_{3,0}$  &  &      \\
            \hline
            CH$_3$CONH$_2$-A & 7$_{2,6,0}$-6$_{1,5,0}$ &  (-4, -3) &  47 \\
                             & 8$_{0,8,0}$-7$_{1,7,0}$ &   &  \\
                             & 8$_{1,8,0}$-7$_{1,7,0}$ & & \\
                             & 8$_{0,8,0}$-7$_{0,7,0}$ &  &   \\
                             & 8$_{1,8,0}$-7$_{0,7,0}$ &  &  \\
            \hline
        \end{tabular}
    \end{center}
    Note: the emission size listed  in Col(3) is defined by the equivalent radius of the area where the emission is stronger than half of the emission at the peak position.
\end{table}

\clearpage

\begin{table}
    \begin{center}
        \caption{The intensity ratio between H$^{13}$CONH$_2$ and HCONH$_2$}
        \label{13C_and_12C}
        \begin{tabular}{cccccc}
            \hline
            \hline
            position & Velocity range & I$_{H^{13}CONH_2}$ & I$_{HCONH_2}$ & ratio & $\tau$ \\
            & km s$^{-1}$ & K km s$^{-1}$ & K km s$^{-1}$ & \% & \\
            \hline
            (0, -60) & 48 75 & 0.44(0.08) & 9.93(0.11) & 22.7(4.0) & 0.51 \\
            (-30, 0) & 55 85 & 0.54(0.05) & 14.28(0.10) & 26.4(2.5) & 0.18 \\
            (-30, -30) & 50 80 & 0.52(0.05) & 15.81(0.10) & 30.7(2.7) & 0.04\\
            (0, 60)& 50 86 & 0.34(0.05) & 12.73(0.07)  & 37.9(6.0) & 0\\
            (30, 60)& 50 85 &  0.53(0.08) & 14.29(0.09) & 27.2(3.9) & 0.11\\
            (-60, 30)& 58 80 & 0.26(0.05) & 8.15(0.08) & 31.3(5.9) & 0\\
            (-30, -60)& 48 78 & 0.67(0.07) & 18.64(0.10) & 27.9(3.1) & 0.06\\
            (0, 90)& 45 85 & 0.21(0.06) & 11.02(0.14) & 51.8(15.6) & 0\\
            (0, -90)& 45 70 & 0.46(0.05) & 14.66(0.07) & 31.9(3.1) & 0\\
            (0, -120)& 48 78 &  0.31(0.6) & 11.11(0.07) & 35.3(7.2) & 0\\
            (60, 60)& 48 78 & 0.36(0.07) & 11.98(0.08) & 33.3(6.2) & 0\\
            (0, 0)  & 55 78 & 1.53(0.17) & 8.96(0.09)  & 5.9(0.7) & 5.30\\
            west Sgr B2 & 60 75 & 0.09(0.01) & 2.7(0.02) & 28.7(1.7) &  \\
            \hline
        \end{tabular}
    \end{center}       
\end{table}

\clearpage

\begin{table}
    \begin{center}
        \caption{Column density of HCONH$_2$ and CH$_3$CONH$_2$}
        \label{columndensity}
        \begin{tabular}{llcccccc}
            \hline
            \hline
            rac & decl & T$_{ex}$  & N$_{HCONH_2}$ &  N$_{CH_3CONH_2}$ & N$_{HNCO}$ & r(CH$_3$CONH$_2$/HCONH$_2$) & r(HCONH$_2$/HNCO) \\
            $\arcsec$ & $\arcsec$ & K & $\times 10^{14} cm^{-2}$ & $\times 10^{13} cm^{-2}$ & $\times 10^{15} cm^{-2}$ & \% & \% \\
            \hline
            -60 & -60 & 14.2$\pm$4.0 &  0.8$\pm$0.2 &  1.0$\pm$0.4 &  1.8$\pm$0.4 &  12.9 &  4.1 \\
30 & 0 & 27.8$\pm$11.9 &  0.9$\pm$0.3 & $\leq$ 0.6 &  2.0$\pm$0.3 & $\leq$ 7.0 &  4.4 \\
-30 & 0 & 26.3$\pm$5.0 &  2.5$\pm$0.4 &  2.3$\pm$0/4  &  4.1$\pm$0.3  & 9.1  &  6.3 \\

0 & 30 & 19.4$\pm$4.5 &  1.4$\pm$0.3 &  1.9$\pm$0.5 &  3.4$\pm$0.4 &  14.3 &  4.0 \\
-30 & 30 & 15.9$\pm$4.9 &  1.2$\pm$0.3 &  2.3$\pm$0.9 &  3.0$\pm$0.6 &  19.4 &  4.0 \\
-30 & -30 & 27.2$\pm$8.8 &  3.1$\pm$0.7 &  4.6$\pm$1.1 &  4.8$\pm$0.6 &  14.8 &  6.5 \\
30 & 30 & 19.7$\pm$11.6 &  1.3$\pm$0.6 &  2.6$\pm$1.5 &  3.5$\pm$1.1 &  19.9 &  3.8 \\
30 & -30 & 33.4$\pm$14.0 &  1.1$\pm$0.3 & $\leq$ 0.6 &  1.9$\pm$0.3 & $\leq$ 5.7 &  5.4 \\
0 & 60 & 13.4$\pm$4.0 &  1.1$\pm$0.4 &  2.2$\pm$1.0 &  3.3$\pm$0.8 &  19.6 &  3.5 \\
60 & 0 & 18.9$\pm$13.1 &  0.3$\pm$0.2 & $\leq$ 0.5 &  1.0$\pm$0.4 & $\leq$ 16.6 &  2.8 \\
-60 & 0 & 16.3$\pm$7.9 &  0.9$\pm$0.4 &  1.5$\pm$0.9 &  2.5$\pm$0.8 &  15.7 &  3.8 \\
0 & -60 & 46.1$\pm$15.8 &  4.6$\pm$0.7 &  7.6$\pm$1.2 & 4.5$\pm$0.4  & 16.7 &  10.2 \\

30 & 60 & 13.1$\pm$3.0 &  1.3$\pm$0.3 &  2.6$\pm$0.9 & 3.4$\pm$0.6  & 19.6  &  3.9 \\

-30 & 60 & 15.5$\pm$6.9 &  0.7$\pm$0.2 &  1.3$\pm$0.8 &  3.1$\pm$0.9 &  19.6 &  2.2 \\
60 & 30 & 12.3$\pm$3.7 &  0.5$\pm$0.2 &  0.8$\pm$0.4 &  1.7$\pm$0.4 &  16.5 &  3.0 \\
-60 & 30 & 12.9$\pm$3.6 &  0.8$\pm$0.3 &  1.8$\pm$0.8 &  2.9$\pm$0.7 &  22.7 &  2.8 \\
60 & -30 & 23.1$\pm$15.1 &  0.4$\pm$0.2 & $\leq$ 0.6 &  1.0$\pm$0.3 & $\leq$ 15.6 &  4.2 \\
-60 & -30 & 15.5$\pm$5.5 &  1.1$\pm$0.3 &  1.4$\pm$0.7 &  2.6$\pm$0.6 &  13.3 &  4.2 \\
30 & -60 & 23.2$\pm$8.6 &  0.8$\pm$0.2 & $\leq$ 0.6 &  1.2$\pm$0.2 & $\leq$ 7.2 &  6.2 \\
-30 & -60 & 24.0$\pm$7.4 &  3.2$\pm$0.7 &  4.8$\pm$1.3 &  3.6$\pm$0.5 &  15.2  &  8.9 \\

-90 & 0 & 14.8$\pm$9.8 &  0.3$\pm$0.2 & $\leq$ 0.4 &  1.7$\pm$0.8 & $\leq$ 11.7 &  2.1 \\
0 & 90 & 9.9$\pm$1.5 &  0.7$\pm$0.1 &  1.2$\pm$0.4 &  2.9$\pm$0.5 &  16.9 &  2.4 \\
0 & -90 & 17.3$\pm$5.3 &  1.8$\pm$0.5 &  2.1$\pm$0.8 &  1.7$\pm$0.3 &  12.1 &  10.1 \\
60 & 90 & 9.5$\pm$1.0 &  0.6$\pm$0.1 & $\leq$ 0.4 &  2.3$\pm$0.3 & $\leq$ 6.7 &  2.6 \\
30 & 90 & 10.8$\pm$3.5 &  0.7$\pm$0.3 &  1.4$\pm$0.8 &  3.0$\pm$0.9 &  19.6 &  2.4 \\
-30 & 90 & 10.0$\pm$2.3 &  0.6$\pm$0.2 &  1.5$\pm$0.7 &  2.4$\pm$0.6 &  25.7 &  2.5 \\
-60 & 90 & 16.7$\pm$18.7 &  0.4$\pm$0.3 & $\leq$ 0.5 &  1.8$\pm$1.3 & $\leq$ 14.2 &  2.0 \\
-90 & 90 & 9.7$\pm$3.5 &  0.3$\pm$0.1 & $\leq$ 0.4 &  1.3$\pm$0.5 & $\leq$ 16.5 &  2.0 \\
60 & -90 & 19.4$\pm$15.6 &  0.6$\pm$0.3 & $\leq$ 0.5 &  1.3$\pm$0.6 & $\leq$ 9.8 &  4.3 \\
30 & -90 & 16.9$\pm$7.6 &  0.7$\pm$0.3 & $\leq$ 0.5 &  1.0$\pm$0.3 & $\leq$ 6.5 &  7.0 \\
-30 & -90 & 15.3$\pm$4.3 &  1.4$\pm$0.4 &  1.9$\pm$0.7 &  2.0$\pm$0.4 &  13.6 &  6.9 \\
-60 & -90 & 14.8$\pm$10.8 &  0.4$\pm$0.3 & $\leq$ 0.5 &  1.3$\pm$0.7 & $\leq$ 12.0 &  3.2 \\
-90 & -90 & 10.2$\pm$3.3 &  0.2$\pm$0.1 & $\leq$ 0.4 &  1.0$\pm$0.3 & $\leq$ 21.5 &  2.1 \\
-90 & 30 & 13.0$\pm$7.5 &  0.5$\pm$0.3 &  0.6$\pm$0.6 &  2.0$\pm$0.9 &  12.6 &  2.5 \\
-90 & 60 & 11.9$\pm$6.2 &  0.4$\pm$0.2 & $\leq$ 0.4 &  1.7$\pm$0.8 & $\leq$ 9.5 &  2.4 \\
-120 & 0 & 15.7$\pm$14.6 &  0.4$\pm$0.3 & $\leq$ 0.5 &  1.8$\pm$1.1 & $\leq$ 14.4 &  2.0 \\
-120 & 30 & 12.3$\pm$7.1 &  0.4$\pm$0.2 & $\leq$ 0.4 &  1.9$\pm$0.9 & $\leq$ 10.8 &  2.2 \\
-120 & 60 & 12.5$\pm$5.9 &  0.3$\pm$0.1 & $\leq$ 0.3 &  1.5$\pm$0.6 & $\leq$ 11.7 &  2.0 \\

            \hline
        \end{tabular}
    \end{center}
\end{table}

\begin{table}
    \begin{center}
        \setcounter{table}{3}
        \caption{Continued.}
        \label{columndensity}
        \begin{tabular}{llcccccc}
            \hline
            \hline
            rac & decl & T$_{ex}$  & N$_{HCONH_2}$ &  N$_{CH_3CONH_2}$ & N$_{HNCO}$ & r(CH$_3$CONH$_2$/HCONH$_2$) & r(HCONH$_2$/HNCO) \\
            $\arcsec$ & $\arcsec$ & K & $\times 10^{14} cm^{-2}$ & $\times 10^{13} cm^{-2}$ & $\times 10^{15} cm^{-2}$ & \% & \% \\
            \hline
            -90 & -30 & 16.6$\pm$12.8 &  0.4$\pm$0.2 & $\leq$ 0.4 &  1.9$\pm$0.9 & $\leq$ 10.8 &  2.0 \\
-90 & -60 & 19.4$\pm$26.0 &  0.4$\pm$0.3 & $\leq$ 0.5 &  1.5$\pm$1.1 & $\leq$ 14.4 &  2.4 \\
0 & 120 & 7.1$\pm$1.4 &  0.3$\pm$0.1 & $\leq$ 0.6 &  1.9$\pm$0.5 & $\leq$ 16.8 &  1.8 \\
30 & 120 & 6.2$\pm$0.2 &  0.3$\pm$0.0 & $\leq$ 1.0 &  2.2$\pm$0.1 & $\leq$ 28.5 &  1.6 \\
60 & 120 & 7.6$\pm$1.1 &  0.2$\pm$0.1 & $\leq$ 0.7 &  1.6$\pm$0.3 & $\leq$ 32.2 &  1.5 \\
-30 & -120 & 14.7$\pm$4.4 &  0.5$\pm$0.1 & $\leq$ 0.4 &  1.2$\pm$0.2 & $\leq$ 8.0 &  4.4 \\
0 & -120 & 13.2$\pm$3.9 &  1.0$\pm$0.3 &  1.4$\pm$0.6 &  1.6$\pm$0.4 &  14.4 &  6.2 \\
30 & -120 & 12.4$\pm$3.5 &  0.7$\pm$0.2 &  0.9$\pm$0.4 &  1.1$\pm$0.3 &  12.4 &  6.5 \\
60 & 60 & 11.4$\pm$2.1 &  1.0$\pm$0.2 &  1.9$\pm$0.6 &  3.0$\pm$0.5 &  19.2 &  3.3 \\
-60 & 60 & 14.5$\pm$6.3 &  0.6$\pm$0.2 &  1.1$\pm$0.7 &  2.3$\pm$0.7 &  19.6 &  2.4 \\
60 & -60 & 15.8$\pm$5.7 &  0.6$\pm$0.2 &  0.6$\pm$0.3 &  1.0$\pm$0.2 &  10.7 &  5.5 \\
60 & -120 & 21.6$\pm$30.1 &  0.5$\pm$0.5 & $\leq$ 0.7 &  1.3$\pm$0.9 & $\leq$ 12.5 &  4.0 \\
0 & -150 & 18.1$\pm$18.3 &  0.8$\pm$0.6 &  1.1$\pm$1.2 &  1.5$\pm$0.9 &  12.9 &  5.4 \\
30 & -150 & 12.1$\pm$6.6 &  0.6$\pm$0.3 &  0.8$\pm$0.7 &  1.1$\pm$0.5 &  13.8 &  5.0 \\
0 & -180 & 14.1$\pm$8.8 &  0.5$\pm$0.3 & $\leq$ 0.6 &  1.7$\pm$0.8 & $\leq$ 11.5 &  3.0 \\
30 & -180 & 12.2$\pm$7.7 &  0.4$\pm$0.2 & $\leq$ 0.4 &  1.0$\pm$0.5 & $\leq$ 11.9 &  3.7 \\
           
            \hline
        \end{tabular}
    \end{center}
\end{table}
Note. Col(1) and Col(2): the equatorial offsets of emission with respect to Sgr B2(N); Col(3): the excitation temperature; Col(4): the column densities of HCONH$_2$; Col(5): the column densities of CH$_3$CONH$_2$; Col(6): the column densities of HNCO; Col(7): the column density ratio of CH$_3$CONH$_2$ to HCONH$_2$.   Col(8): the column density ratio of HCONH$_2$ to HNCO.

\clearpage


\appendix

\renewcommand{\thetable}{A\arabic{table}}
\renewcommand{\thefigure}{A\arabic{figure}}
\setcounter{table}{0}
\startlongtable

\begin{deluxetable*}{llccccccccc}
    \tabletypesize{\scriptsize}
    \tablewidth{100pt} 
    \tablecaption{The integrated intensities, the first moment (v$_{m1}$, the intensity-weighted velocity) and second moment ($\sigma_{m2},$the intensity-weighted velocity dispersion) of the transitions of HCONH$_2$. \label{intensity_hconh2_1}}
    \tablehead{
    \colhead{rac} & \colhead{dec} & \multicolumn{3}{c}{4$_{3,2}$-3$_{3,1}$ and 4$_{3,1}$-3$_{3,0}$} & \multicolumn{3}{c}{1$_{1,1}$-0$_{0,0}$} & \multicolumn{3}{c}{4$_{0,4}$-3$_{0,3}$}\\
    \cline{3-11}
    \colhead{} & \colhead{} &
    \colhead{intensity} & \colhead{v$_{m1}$} & \colhead{$\sigma_{m2}$} & \colhead{intensity} & \colhead{v$_{m1}$} & \colhead{$\sigma_{m2}$} &\colhead{intensity} & \colhead{v$_{m1}$} & \colhead{$\sigma_{m2}$} \\
    \colhead{\arcsec} & \colhead{\arcsec} &
    \colhead{K km s$^{-1}$} & \colhead{km s$^{-1}$} & \colhead{km s$^{-1}$} & \colhead{K km s$^{-1}$} & \colhead{km s$^{-1}$} & \colhead{km s$^{-1}$} &\colhead{K km s$^{-1}$} & \colhead{km s$^{-1}$} & \colhead{km s$^{-1}$}
    }
    \colnumbers
    \startdata 
            -60.0 & -60.0 & 0.9$\pm$0.1 & 57.4 & 18.4 & 2.2$\pm$0.1 & 62.3 & 17.1 & 7.7$\pm$0.1 & 63.6 & 21.1  \\
            30.0 & 0.0 & 1.7$\pm$0.1 & 67.2 & 24.3 & 0.8$\pm$0.0 & 65.3 & 14.8 & 4.2$\pm$0.1 & 69.6 & 21.1  \\
            -30.0 & 0.0 & 5.2$\pm$0.1 & 63.8 & 15.9 & 3.0$\pm$0.1 & 69.2 & 16.8 & 14.9$\pm$0.1 & 69.9 & 16.0  \\
            0.0 & 30.0 & 2.7$\pm$0.1 & 64.7 & 18.4 & 3.7$\pm$0.1 & 70.4 & 20.6 & 10.8$\pm$0.1 & 69.5 & 17.9  \\
            0.0 & -30.0 & 6.7$\pm$0.1 & 59.3 & 14.5 & 1.3$\pm$0.1 & 64.6 & 15.8 & 0.0$\pm$0.0 & 61.6 & 29.2  \\
            -30.0 & 30.0 & 1.3$\pm$0.1 & 63.0 & 14.5 & 3.7$\pm$0.1 & 68.0 & 17.9 & 11.5$\pm$0.1 & 69.2 & 14.6  \\
            -30.0 & -30.0 & 5.5$\pm$0.1 & 58.8 & 23.4 & 3.7$\pm$0.1 & 65.4 & 19.9 & 18.2$\pm$0.1 & 64.5 & 21.6  \\
            30.0 & 30.0 & 1.2$\pm$0.1 & 64.8 & 18.4 & 3.9$\pm$0.0 & 68.6 & 22.2 & 9.6$\pm$0.1 & 69.6 & 23.1  \\
            30.0 & -30.0 & 1.9$\pm$0.0 & 61.7 & 19.5 & 0.7$\pm$0.1 & 62.1 & 16.4 & 4.0$\pm$0.1 & 65.2 & 19.0  \\
            0.0 & 60.0 & 1.0$\pm$0.1 & 64.1 & 17.2 & 4.1$\pm$0.1 & 67.8 & 21.4 & 13.1$\pm$0.1 & 68.6 & 20.6  \\
            60.0 & 0.0 & 0.4$\pm$0.1 & 56.0 & 13.0 & 1.0$\pm$0.1 & 62.3 & 18.1 & 1.6$\pm$0.1 & 60.5 & 13.8  \\
            -60.0 & 0.0 & 0.8$\pm$0.1 & 61.3 & 15.9 & 2.7$\pm$0.1 & 66.2 & 20.7 & 8.9$\pm$0.1 & 67.8 & 20.6  \\
            0.0 & -60.0 & 7.1$\pm$0.1 & 58.9 & 15.9 & 2.1$\pm$0.1 & 62.4 & 18.0 & 10.7$\pm$0.1 & 59.5 & 15.3  \\
            30.0 & 60.0 & 1.2$\pm$0.1 & 65.2 & 22.5 & 4.3$\pm$0.1 & 66.7 & 19.7 & 15.0$\pm$0.1 & 68.2 & 24.4  \\
            -30.0 & 60.0 & 0.3$\pm$0.1 & 64.7 & 13.0 & 2.5$\pm$0.1 & 67.2 & 18.0 & 7.1$\pm$0.1 & 66.6 & 16.6  \\
            60.0 & 30.0 & 0.6$\pm$0.1 & 60.0 & 17.2 & 2.5$\pm$0.1 & 62.7 & 19.1 & 5.0$\pm$0.1 & 63.7 & 16.6  \\
            -60.0 & 30.0 & 0.7$\pm$0.1 & 60.8 & 20.5 & 3.2$\pm$0.1 & 68.1 & 17.6 & 9.4$\pm$0.1 & 69.4 & 20.1  \\
            60.0 & -30.0 & $\leq$ 0.2 & 57.3 & 22.5 & 0.7$\pm$0.1 & 57.2 & 17.7 & 3.0$\pm$0.1 & 58.4 & 17.9   \\
            -60.0 & -30.0 & 1.3$\pm$0.1 & 56.5 & 21.5 & 3.0$\pm$0.1 & 63.0 & 20.2 & 10.3$\pm$0.1 & 63.4 & 22.6  \\
            30.0 & -60.0 & 1.7$\pm$0.1 & 58.2 & 21.5 & 1.3$\pm$0.1 & 60.6 & 19.1 & 4.0$\pm$0.1 & 59.2 & 13.0  \\
            -30.0 & -60.0 & 5.1$\pm$0.1 & 56.0 & 20.5 & 4.0$\pm$0.1 & 63.4 & 19.8 & 20.5$\pm$0.1 & 61.1 & 22.6  \\
            -90.0 & 0.0 & $\leq$ 0.2 & 58.2 & 0.0 & 1.3$\pm$0.1 & 63.9 & 22.4 & 4.0$\pm$0.1 & 68.2 & 19.6   \\
            0.0 & 90.0 & 0.4$\pm$0.0 & 58.7 & 17.2 & 3.0$\pm$0.1 & 62.1 & 20.9 & 11.1$\pm$0.1 & 61.2 & 19.6  \\
            0.0 & -90.0 & 2.6$\pm$0.1 & 55.0 & 17.2 & 3.0$\pm$0.1 & 58.8 & 17.3 & 15.6$\pm$0.1 & 56.9 & 21.1  \\
            60.0 & 90.0 & $\leq$ 0.2 & 60.1 & 22.5 & 2.3$\pm$0.1 & 60.4 & 22.1 & 9.8$\pm$0.1 & 61.7 & 20.6   \\
            30.0 & 90.0 & $\leq$ 0.2 & 53.5 & 18.4 & 3.1$\pm$0.1 & 63.2 & 22.4 & 10.3$\pm$0.1 & 60.8 & 19.6   \\
            -30.0 & 90.0 & $\leq$ 0.2 & 55.9 & 18.4 & 2.0$\pm$0.1 & 61.8 & 18.6 & 8.8$\pm$0.1 & 61.3 & 13.0   \\
            -60.0 & 90.0 & $\leq$ 0.2 & 63.6 & 0.0 & 1.4$\pm$0.1 & 62.1 & 19.7 & 3.3$\pm$0.1 & 62.8 & 17.3   \\
            -90.0 & 90.0 & $\leq$ 0.2 & 44.1 & 0.0 & 1.5$\pm$0.1 & 60.4 & 20.4 & 3.9$\pm$0.1 & 62.3 & 17.9   \\
            60.0 & -90.0 & 0.4$\pm$0.1 & 62.8 & 18.4 & 1.6$\pm$0.1 & 60.7 & 15.1 & 3.7$\pm$0.1 & 62.2 & 17.3  \\
            30.0 & -90.0 & 1.0$\pm$0.1 & 55.2 & 20.5 & 1.4$\pm$0.1 & 61.0 & 17.3 & 5.9$\pm$0.1 & 57.8 & 13.8  \\
            -30.0 & -90.0 & 1.7$\pm$0.1 & 55.6 & 22.5 & 3.0$\pm$0.1 & 61.3 & 18.7 & 13.4$\pm$0.1 & 59.5 & 20.1  \\
            -60.0 & -90.0 & $\leq$ 0.2 & 63.3 & 18.4 & 1.2$\pm$0.1 & 62.1 & 20.3 & 4.4$\pm$0.1 & 60.4 & 19.6   \\
            -90.0 & -90.0 & $\leq$ 0.2 & 56.6 & 0.0 & 0.9$\pm$0.1 & 64.3 & 14.0 & 3.0$\pm$0.1 & 59.5 & 20.6   \\
            -90.0 & 30.0 & $\leq$ 0.2 & 67.7 & 14.5 & 2.0$\pm$0.1 & 72.7 & 17.4 & 6.2$\pm$0.1 & 71.9 & 16.6   \\
            -90.0 & 60.0 & $\leq$ 0.2 & 63.4 & 14.5 & 1.5$\pm$0.1 & 66.1 & 18.7 & 5.5$\pm$0.1 & 69.1 & 13.8   \\
            -120.0 & 0.0 & $\leq$ 0.3 & -78.5 & 0.0 & 1.2$\pm$0.1 & 62.1 & 24.7 & 3.7$\pm$0.1 & 64.0 & 28.8   \\
            -120.0 & 30.0 & $\leq$ 0.2 & 58.4 & 27.6 & 2.0$\pm$0.1 & 62.8 & 22.7 & 4.9$\pm$0.1 & 63.7 & 24.0   \\
            -120.0 & 60.0 & $\leq$ 0.2 & 53.7 & 19.5 & 1.5$\pm$0.1 & 61.1 & 19.4 & 4.0$\pm$0.1 & 60.3 & 22.1   \\
            -90.0 & -30.0 & $\leq$ 0.2 & 57.1 & 13.0 & 1.4$\pm$0.1 & 65.5 & 20.2 & 3.7$\pm$0.1 & 64.0 & 16.6   \\
            -120.0 & -30.0 & $\leq$ 0.2 & 54.5 & 0.0 & 0.9$\pm$0.1 & 68.3 & 24.5 & 3.0$\pm$0.1 & 62.4 & 31.3   \\
            -90.0 & -60.0 & $\leq$ 0.2 & 54.6 & 18.4 & 0.9$\pm$0.1 & 64.0 & 16.2 & 2.8$\pm$0.1 & 62.8 & 9.2   \\
            -120.0 & -60.0 & $\leq$ 0.2 & 60.8 & 22.5 & $\leq$ 0.2 & 59.1 & 24.2 & 1.8$\pm$0.1 & 61.9 & 19.6   \\
            0.0 & 120.0 & $\leq$ 0.2 & 61.6 & 27.6 & 1.0$\pm$0.1 & 61.0 & 19.6 & 5.6$\pm$0.1 & 49.5 & 38.0   \\
            30.0 & 120.0 & $\leq$ 0.3 & 51.1 & 27.6 & 1.1$\pm$0.2 & 58.8 & 33.4 & 5.9$\pm$0.1 & 49.8 & 38.0   \\
            60.0 & 120.0 & $\leq$ 0.3 & 55.0 & 47.7 & 1.3$\pm$0.1 & 61.5 & 23.5 & 3.9$\pm$0.1 & 59.2 & 24.0   \\
            90.0 & 90.0 & $\leq$ 0.2 & 56.5 & 42.6 & 1.0$\pm$0.1 & 57.8 & 15.4 & 3.4$\pm$0.1 & 57.7 & 13.8   \\
            90.0 & 60.0 & $\leq$ 0.2 & 0.0 & 0.0 & 1.7$\pm$0.1 & 58.9 & 28.2 & 0.0$\pm$0.0 & 53.5 & 25.3   \\
            90.0 & 120.0 & $\leq$ 0.2 & 49.6 & 0.0 & $\leq$ 0.2 & 60.3 & 13.1 & 2.2$\pm$0.1 & 61.5 & 16.0   \\
            -30.0 & -120.0 & 0.6$\pm$0.1 & 54.8 & 22.5 & 1.3$\pm$0.1 & 58.9 & 20.6 & 5.8$\pm$0.1 & 60.0 & 21.1  \\
            0.0 & -120.0 & 1.2$\pm$0.1 & 51.9 & 17.2 & 2.4$\pm$0.1 & 58.4 & 18.6 & 12.0$\pm$0.1 & 57.2 & 16.0  \\
            30.0 & -120.0 & 0.8$\pm$0.1 & 50.1 & 18.4 & 1.5$\pm$0.1 & 56.1 & 11.6 & 8.4$\pm$0.1 & 55.7 & 10.3  \\
            60.0 & 60.0 & 0.9$\pm$0.1 & 61.5 & 17.2 & 3.4$\pm$0.1 & 63.8 & 19.8 & 13.2$\pm$0.1 & 61.8 & 22.1  \\
            0.0 & 0.0 & 22.3$\pm$0.2 & 61.7 & 15.9 & 0.0$\pm$0.0 & 67.8 & 6.7 & 0.0$\pm$0.0 & 61.5 & 0.0  \\
            -60.0 & 60.0 & $\leq$ 0.3 & 64.9 & 25.2 & 2.3$\pm$0.1 & 67.8 & 18.7 & 5.9$\pm$0.1 & 68.8 & 13.0   \\
            60.0 & -60.0 & 0.8$\pm$0.1 & 56.7 & 22.5 & 1.7$\pm$0.1 & 57.4 & 17.5 & 4.5$\pm$0.1 & 58.0 & 17.9  \\
            30.0 & 150.0 & $\leq$ 0.2 & 63.4 & 0.0 & $\leq$ 0.3 & 61.7 & 14.6 & 0.8$\pm$0.1 & 60.3 & 11.3   \\
            60.0 & -120.0 & $\leq$ 0.2 & 53.9 & 26.0 & 1.5$\pm$0.1 & 58.3 & 18.6 & 3.6$\pm$0.1 & 57.4 & 16.0   \\
            0.0 & -150.0 & $\leq$ 0.2 & 58.6 & 20.5 & 1.4$\pm$0.1 & 60.3 & 16.1 & 7.3$\pm$0.1 & 60.1 & 18.5   \\
            30.0 & -150.0 & $\leq$ 0.2 & 52.4 & 18.4 & 1.0$\pm$0.1 & 56.1 & 12.9 & 6.7$\pm$0.1 & 55.6 & 15.3   \\
            0.0 & -180.0 & $\leq$ 0.2 & 55.6 & 22.5 & 1.8$\pm$0.1 & 55.2 & 16.5 & 5.9$\pm$0.1 & 55.5 & 14.6   \\
            30.0 & -180.0 & $\leq$ 0.2 & 50.7 & 11.3 & 1.0$\pm$0.1 & 52.6 & 14.2 & 4.4$\pm$0.1 & 51.7 & 14.6   \\
            0.0 & -48.0 & 7.2$\pm$0.1 & 59.6 & 17.2 & 0.9$\pm$0.1 & 63.6 & 18.4 & 0.0$\pm$0.0 & 54.4 & 33.3  \\
    \enddata
    \end{deluxetable*}

\clearpage
\startlongtable
\begin{deluxetable*}{llccccccccc}
    \tabletypesize{\scriptsize}
    \tablewidth{100pt} 
    \tablecaption{The integrated intensities, the first moment (v$_{m1}$, the intensity-weighted velocity) and second moment ($\sigma_{m2},$the intensity-weighted velocity dispersion) of the transitions of HCONH$_2$. \label{intensity_hconh2_2}}
    \tablehead{
    \colhead{rac} & \colhead{dec} & \multicolumn{3}{c}{4$_{1,3}$-3$_{1,2}$} & \multicolumn{3}{c}{4$_{2,2}$-3$_{2,1}$} & \multicolumn{3}{c}{4$_{2,3}$-3$_{2,2}$ }\\
    \cline{3-11}
    \colhead{} & \colhead{} &
    \colhead{intensity} & \colhead{v$_{m1}$} & \colhead{$\sigma_{m2}$} & \colhead{intensity} & \colhead{v$_{m1}$} & \colhead{$\sigma_{m2}$} &\colhead{intensity} & \colhead{v$_{m1}$} & \colhead{$\sigma_{m2}$} \\
    \colhead{\arcsec} & \colhead{\arcsec} &
    \colhead{K km s$^{-1}$} & \colhead{km s$^{-1}$} & \colhead{km s$^{-1}$} & \colhead{K km s$^{-1}$} & \colhead{km s$^{-1}$} & \colhead{km s$^{-1}$} &\colhead{K km s$^{-1}$} & \colhead{km s$^{-1}$} & \colhead{km s$^{-1}$}
    }
    \colnumbers
    \startdata 
        -60.0 & -60.0 & 9.7$\pm$0.1 & 63.1 & 22.2 & 2.9$\pm$0.1 & 61.7 & 14.5 & 3.0$\pm$0.1 & 62.9 & 15.9 \\
30.0 & 0.0 & 6.6$\pm$0.1 & 69.7 & 23.1 & 3.0$\pm$0.1 & 69.8 & 23.4 & 2.9$\pm$0.1 & 70.0 & 24.3 \\
-30.0 & 0.0 & 19.8$\pm$0.1 & 69.7 & 19.4 & 9.2$\pm$0.1 & 68.5 & 13.0 & 9.0$\pm$0.1 & 68.6 & 19.5 \\
0.0 & 30.0 & 13.4$\pm$0.1 & 69.0 & 22.2 & 5.2$\pm$0.1 & 68.2 & 19.4 & 5.1$\pm$0.1 & 68.5 & 19.5 \\
0.0 & -30.0 & 13.7$\pm$0.1 & 63.5 & 25.5 & 10.0$\pm$0.1 & 63.0 & 24.3 & 9.9$\pm$0.1 & 63.1 & 21.6 \\
-30.0 & 30.0 & 14.0$\pm$0.1 & 68.4 & 18.3 & 4.6$\pm$0.1 & 67.9 & 15.9 & 4.7$\pm$0.0 & 68.4 & 17.2 \\
-30.0 & -30.0 & 25.0$\pm$0.1 & 64.4 & 24.7 & 11.8$\pm$0.1 & 62.9 & 17.1 & 12.3$\pm$0.1 & 63.9 & 22.5 \\
30.0 & 30.0 & 12.8$\pm$0.1 & 68.9 & 22.6 & 4.5$\pm$0.1 & 67.0 & 24.3 & 4.3$\pm$0.1 & 68.8 & 21.6 \\
30.0 & -30.0 & 6.0$\pm$0.1 & 65.1 & 20.8 & 3.2$\pm$0.1 & 65.8 & 14.5 & 3.5$\pm$0.1 & 66.4 & 17.2 \\
0.0 & 60.0 & 14.3$\pm$0.1 & 68.8 & 21.3 & 4.2$\pm$0.1 & 68.5 & 17.1 & 4.1$\pm$0.1 & 69.5 & 24.3 \\
60.0 & 0.0 & 3.3$\pm$0.1 & 63.6 & 28.1 & 1.0$\pm$0.1 & 64.9 & 13.0 & 0.9$\pm$0.1 & 66.9 & 26.0 \\
-60.0 & 0.0 & 10.9$\pm$0.1 & 67.4 & 21.3 & 3.1$\pm$0.1 & 67.2 & 20.5 & 3.1$\pm$0.1 & 67.3 & 24.3 \\
0.0 & -60.0 & 17.0$\pm$0.1 & 60.4 & 16.6 & 10.2$\pm$0.1 & 61.4 & 17.1 & 9.9$\pm$0.1 & 61.4 & 14.5 \\
30.0 & 60.0 & 17.1$\pm$0.1 & 68.3 & 20.3 & 5.0$\pm$0.1 & 67.4 & 20.5 & 5.1$\pm$0.1 & 69.0 & 19.5 \\
-30.0 & 60.0 & 8.1$\pm$0.1 & 66.8 & 18.8 & 2.3$\pm$0.1 & 67.0 & 18.3 & 2.2$\pm$0.1 & 66.4 & 20.6 \\
60.0 & 30.0 & 6.6$\pm$0.1 & 63.7 & 20.3 & 1.9$\pm$0.0 & 63.7 & 17.1 & 1.8$\pm$0.1 & 64.2 & 19.5 \\
-60.0 & 30.0 & 11.3$\pm$0.1 & 68.6 & 20.3 & 3.2$\pm$0.1 & 68.0 & 11.2 & 3.0$\pm$0.1 & 68.0 & 20.6 \\
60.0 & -30.0 & 3.9$\pm$0.1 & 58.0 & 20.3 & 1.3$\pm$0.1 & 59.2 & 17.1 & 1.4$\pm$0.1 & 60.2 & 22.5 \\
-60.0 & -30.0 & 12.7$\pm$0.1 & 63.4 & 22.2 & 4.1$\pm$0.1 & 62.2 & 20.5 & 4.0$\pm$0.1 & 63.0 & 19.5 \\
30.0 & -60.0 & 6.5$\pm$0.1 & 59.8 & 21.3 & 2.8$\pm$0.1 & 60.9 & 18.3 & 2.6$\pm$0.1 & 60.7 & 19.5 \\
-30.0 & -60.0 & 26.4$\pm$0.1 & 61.4 & 21.3 & 12.0$\pm$0.1 & 60.2 & 23.4 & 12.1$\pm$0.1 & 61.0 & 23.4 \\
-90.0 & 0.0 & 4.2$\pm$0.1 & 67.1 & 20.3 & 1.1$\pm$0.1 & 67.0 & 14.5 & 0.9$\pm$0.1 & 66.3 & 20.6 \\
0.0 & 90.0 & 10.8$\pm$0.2 & 61.3 & 20.8 & 2.5$\pm$0.1 & 59.4 & 20.5 & 2.3$\pm$0.1 & 60.3 & 21.6 \\
0.0 & -90.0 & 19.4$\pm$0.1 & 57.8 & 17.2 & 6.9$\pm$0.1 & 58.0 & 19.4 & 6.8$\pm$0.1 & 58.2 & 23.4 \\
60.0 & 90.0 & 8.2$\pm$0.2 & 60.6 & 24.3 & 2.1$\pm$0.1 & 60.3 & 23.4 & 2.0$\pm$0.1 & 62.2 & 20.6 \\
30.0 & 90.0 & 10.7$\pm$0.2 & 60.2 & 27.0 & 2.4$\pm$0.1 & 61.0 & 19.4 & 2.3$\pm$0.1 & 61.5 & 24.3 \\
-30.0 & 90.0 & 7.6$\pm$0.1 & 61.5 & 19.9 & 1.9$\pm$0.1 & 60.5 & 17.1 & 1.8$\pm$0.1 & 61.8 & 13.0 \\
-60.0 & 90.0 & 4.2$\pm$0.1 & 62.4 & 19.4 & 0.9$\pm$0.1 & 63.5 & 22.5 & 0.8$\pm$0.1 & 62.6 & 22.5 \\
-90.0 & 90.0 & 3.7$\pm$0.1 & 60.5 & 17.8 & 0.7$\pm$0.1 & 63.1 & 15.9 & 0.7$\pm$0.1 & 61.3 & 14.5 \\
60.0 & -90.0 & 6.1$\pm$0.1 & 63.1 & 20.3 & 1.9$\pm$0.1 & 63.5 & 20.5 & 1.5$\pm$0.1 & 63.7 & 14.5 \\
30.0 & -90.0 & 8.7$\pm$0.1 & 58.6 & 19.9 & 2.6$\pm$0.1 & 58.9 & 19.4 & 2.6$\pm$0.1 & 59.3 & 20.6 \\
-30.0 & -90.0 & 16.2$\pm$0.1 & 60.4 & 21.3 & 5.6$\pm$0.1 & 59.8 & 19.4 & 5.3$\pm$0.1 & 59.7 & 19.5 \\
-60.0 & -90.0 & 5.2$\pm$0.1 & 63.5 & 21.7 & 1.2$\pm$0.1 & 63.7 & 9.2 & 1.2$\pm$0.1 & 61.5 & 17.2 \\
-90.0 & -90.0 & 2.7$\pm$0.1 & 65.2 & 19.4 & 0.5$\pm$0.1 & 63.2 & 18.3 & 0.6$\pm$0.1 & 62.5 & 20.6 \\
-90.0 & 30.0 & 6.3$\pm$0.1 & 71.3 & 20.8 & 1.3$\pm$0.1 & 71.0 & 9.2 & 1.3$\pm$0.1 & 71.2 & 17.2 \\
-90.0 & 60.0 & 5.7$\pm$0.1 & 68.4 & 18.3 & 1.2$\pm$0.1 & 69.0 & 9.2 & 1.0$\pm$0.1 & 69.4 & 11.3 \\
-120.0 & 0.0 & 4.0$\pm$0.1 & 64.3 & 27.7 & 0.8$\pm$0.1 & 66.1 & 26.7 & 0.7$\pm$0.1 & 64.6 & 24.3 \\
-120.0 & 30.0 & 5.1$\pm$0.1 & 63.6 & 21.3 & 1.2$\pm$0.1 & 63.2 & 23.4 & 0.9$\pm$0.1 & 67.4 & 25.2 \\
-120.0 & 60.0 & 3.7$\pm$0.1 & 62.0 & 19.9 & 0.8$\pm$0.1 & 62.5 & 14.5 & 0.7$\pm$0.1 & 61.3 & 23.4 \\
-90.0 & -30.0 & 4.2$\pm$0.1 & 64.0 & 19.9 & 1.1$\pm$0.1 & 64.5 & 11.2 & 1.0$\pm$0.1 & 60.7 & 20.6 \\
-120.0 & -30.0 & 2.8$\pm$0.1 & 63.7 & 30.1 & $\leq$ 0.2 & 67.0 & 15.9 & $\leq$ 0.3 & 68.8 & 9.2 \\
-90.0 & -60.0 & 3.5$\pm$0.1 & 63.1 & 17.8 & 0.8$\pm$0.1 & 64.3 & 14.5 & 0.6$\pm$0.1 & 62.9 & 11.3 \\
-120.0 & -60.0 & 2.2$\pm$0.1 & 63.0 & 20.3 & $\leq$ 0.2 & 56.5 & 13.0 & $\leq$ 0.2 & 63.9 & 24.3 \\
0.0 & 120.0 & 4.0$\pm$0.1 & 55.0 & 28.1 & 0.6$\pm$0.1 & 60.2 & 18.3 & 0.7$\pm$0.1 & 59.2 & 15.9 \\
30.0 & 120.0 & 3.3$\pm$0.1 & 61.7 & 18.8 & 0.7$\pm$0.1 & 46.9 & 29.0 & 0.7$\pm$0.1 & 58.7 & 17.2 \\
60.0 & 120.0 & 2.8$\pm$0.1 & 58.7 & 20.8 & 0.4$\pm$0.1 & 49.4 & 0.0 & 0.6$\pm$0.1 & 57.3 & 13.0 \\
90.0 & 90.0 & 3.9$\pm$0.1 & 59.4 & 21.7 & 0.7$\pm$0.1 & 58.1 & 9.2 & 0.7$\pm$0.1 & 60.8 & 20.6 \\
90.0 & 60.0 & 5.9$\pm$0.1 & 57.7 & 32.0 & 1.2$\pm$0.1 & 57.1 & 28.3 & 1.3$\pm$0.1 & 59.1 & 29.8 \\
90.0 & 120.0 & 2.0$\pm$0.1 & 61.4 & 11.7 & $\leq$ 0.2 & 62.2 & 13.0 & 0.4$\pm$0.1 & 61.3 & 0.0 \\
-30.0 & -120.0 & 6.0$\pm$0.1 & 60.0 & 18.8 & 1.9$\pm$0.1 & 58.4 & 18.3 & 1.5$\pm$0.1 & 58.2 & 17.2 \\
0.0 & -120.0 & 12.6$\pm$0.1 & 57.9 & 16.0 & 3.1$\pm$0.1 & 56.8 & 15.9 & 3.3$\pm$0.1 & 57.0 & 15.9 \\
30.0 & -120.0 & 9.0$\pm$0.1 & 56.1 & 11.7 & 2.3$\pm$0.1 & 55.3 & 0.0 & 2.2$\pm$0.1 & 55.7 & 13.0 \\
60.0 & 60.0 & 13.4$\pm$0.1 & 64.0 & 22.9 & 3.7$\pm$0.0 & 63.9 & 21.5 & 3.4$\pm$0.1 & 64.5 & 14.5 \\
0.0 & 0.0 & 15.4$\pm$0.1 & 64.7 & 13.7 & 18.4$\pm$0.2 & 65.6 & 11.2 & 17.0$\pm$0.1 & 65.2 & 11.3 \\
-60.0 & 60.0 & 7.0$\pm$0.1 & 67.9 & 19.1 & 2.1$\pm$0.0 & 67.8 & 20.5 & 1.6$\pm$0.1 & 69.8 & 13.0 \\
60.0 & -60.0 & 7.0$\pm$0.1 & 59.2 & 20.1 & 2.3$\pm$0.0 & 60.9 & 19.4 & 2.0$\pm$0.1 & 59.7 & 11.3 \\
30.0 & 150.0 & 1.3$\pm$0.1 & 61.7 & 30.4 & $\leq$ 0.2 & 60.0 & 0.0 &$\leq$ 0.2 & -72.2 & 0.0 \\
60.0 & -120.0 & 5.5$\pm$0.1 & 57.3 & 23.5 & 1.4$\pm$0.1 & 58.7 & 11.2 & 1.2$\pm$0.1 & 57.7 & 18.4 \\
0.0 & -150.0 & 8.5$\pm$0.1 & 62.2 & 23.1 & 2.0$\pm$0.1 & 60.5 & 18.3 & 1.8$\pm$0.1 & 60.8 & 9.2 \\
30.0 & -150.0 & 6.9$\pm$0.1 & 56.1 & 14.7 & 1.6$\pm$0.1 & 56.2 & 9.2 & 1.4$\pm$0.1 & 56.0 & 15.9 \\
0.0 & -180.0 & 6.3$\pm$0.1 & 55.7 & 19.4 & 1.3$\pm$0.1 & 55.2 & 19.4 & 1.6$\pm$0.1 & 56.3 & 22.5 \\
30.0 & -180.0 & 4.6$\pm$0.1 & 52.3 & 18.3 & 0.9$\pm$0.1 & 52.4 & 13.0 & 0.9$\pm$0.1 & 52.3 & 13.0 \\
0.0 & -48.0 & 9.4$\pm$0.2 & 61.5 & 28.4 & 8.6$\pm$0.1 & 61.7 & 18.3 & 8.6$\pm$0.1 & 61.5 & 13.0 
\enddata
\end{deluxetable*}

\clearpage
\setcounter{figure}{0}
\begin{figure}
    \centering
    \includegraphics[width=0.9\textwidth]{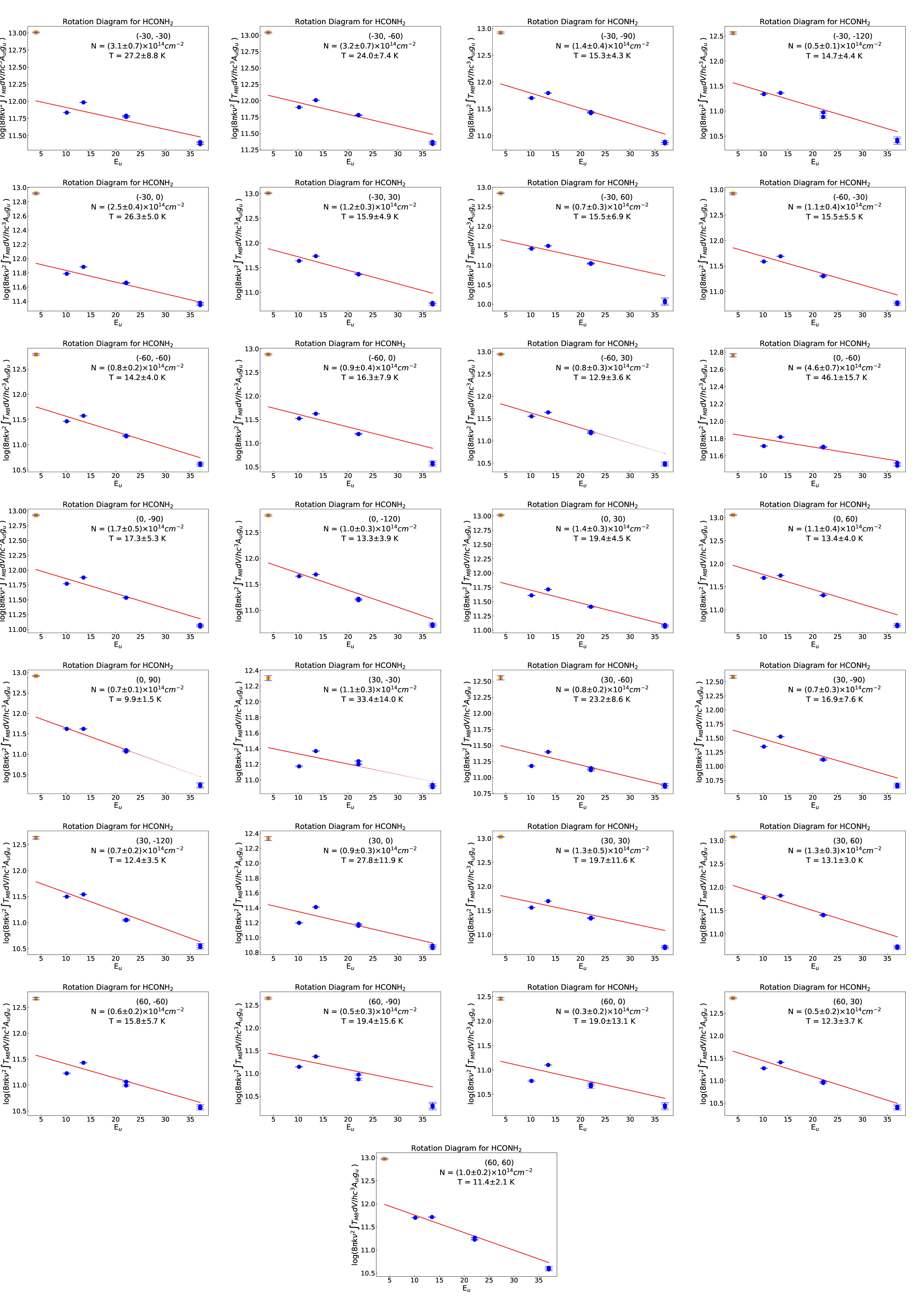}
    \caption{The rotation diagrams of HCONH$_{2}$ of positions with HCONH2$_2$ 4$_3$-3$_2$ stronger than 3 $\sigma$. The error bar is given at 1 $\sigma$ level. HCONH$_2$ 1$_{1,1}$-0$_{0,0}$ is labeled as an orange square, while other transitions are labeled as blue circles.}
    \label{ratotation_diagram}
\end{figure}

\clearpage

\begin{figure}
    \centering
    \includegraphics[width=0.9\textwidth]{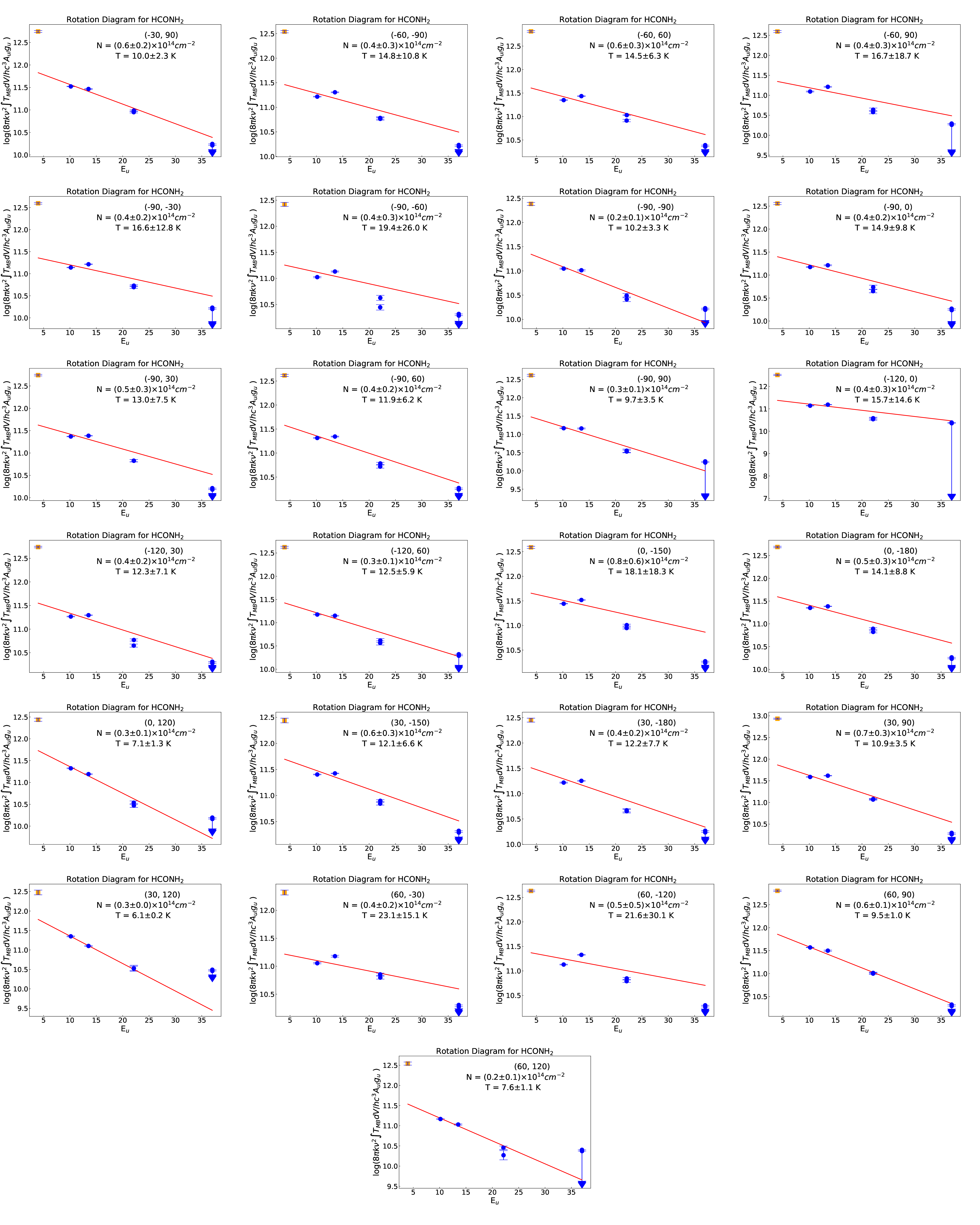}

    \caption{The rotation diagrams of HCONH$_{2}$ of positions where HCONH$_2$ 4$_3$-3$_2$ is not detected. The error bar is given at 1 $\sigma$ level. The downward arrow represents the upper limit of intensity. HCONH$_2$ 1$_{1,1}$-0$_{0,0}$ is labeled as an orange square, while other transitions are labeled as blue circles.}
    \label{ratotation_diagram_3sigma}
\end{figure}

\clearpage

\bibliography{hconh2.bib}{}
\bibliographystyle{aasjournal}

\end{document}